\documentclass[copyright,creativecommons]{eptcs}
\providecommand{\event}{NCMA 2022} % Name of the event you are submitting to

\usepackage{iftex}

\ifpdf
\usepackage{underscore}         % Only needed if you use pdflatex.
\usepackage[T1]{fontenc}        % Recommended with pdflatex
\else
\usepackage{breakurl}           % Not needed if you use pdflatex only.
\fi

\usepackage{graphicx}
\usepackage[english=american]{csquotes}
\usepackage{graphics}
\usepackage[ruled,linesnumbered]{algorithm2e}
\usepackage{amsmath}
\usepackage{amsthm}
\usepackage{tabularx}
\usepackage{environ}
\usepackage{amssymb}
\usepackage{todonotes}
\usepackage{tikz}
\usetikzlibrary{backgrounds}
\usepackage{caption}
\usepackage{subcaption}
\usetikzlibrary{positioning}
\usepgflibrary{arrows}
\usetikzlibrary{arrows,decorations.pathreplacing,calligraphy}
\usepackage{float}
\usepackage[bibliography=common]{apxproof}

\usepackage{thmtools} 
\usepackage{thm-restate}

\usepackage{mathtools}
\usepackage[super]{nth}
\usepackage{todonotes}
\usepackage{hyperref}
\usepackage[capitalise,nameinlink]{cleveref}
\crefname{problem}{Problem}{Problems}

\AtEndEnvironment{proof}{\phantom{}\qed}

\makeatletter
\newcommand{\problemtitle}[1]{\gdef\@problemtitle{#1}}% Store problem title
\newcommand{\probleminput}[1]{\gdef\@probleminput{#1}}% Store problem input
\newcommand{\problemquestion}[1]{\gdef\@problemquestion{#1}}% Store problem question
\NewEnviron{problemdescription}{
	\problemtitle{}\probleminput{}\problemquestion{}% Default input is empty
	\BODY% Parse input
	\par\addvspace{.5\baselineskip}
	\noindent
	\begin{tabularx}{\textwidth}{@{\hspace{\parindent}} l X c}
		\hline
		\multicolumn{2}{@{\hspace{\parindent}}l}{\@problemtitle} \\% Title
		\textbf{Input:} & \@probleminput \\% Input
		\textbf{Question:} & \@problemquestion\\% Question
		\hline
	\end{tabularx}
	\par\addvspace{.5\baselineskip}
}
\makeatother

\def\ta{\mathtt{a}}
\def\tb{\mathtt{b}}
\def\tc{\mathtt{c}}
\def\td{\mathtt{d}}

\def\nth#1{#1$^{\text{th}}$}

\def\N{\mathbb{N}}

\DeclareMathOperator{\al}{alph}
\DeclareMathOperator{\ar}{ar}

\DeclareMathOperator{\bigO}{O}

\DeclareMathOperator{\DFA}{DFA}

\DeclareMathOperator{\ETH}{\textsf{ETH}}

\DeclareMathOperator{\letters}{\al}

\DeclareMathOperator{\mas}{MAS}
\DeclareMathOperator{\masExtend}{masExt}

\DeclareMathOperator{\nz}{\mathsf{nz}}
\DeclareMathOperator{\OV}{\textsf{OV}}
\DeclareMathOperator{\OVH}{\textsf{OVH}}
\DeclareMathOperator{\pmas}{p-MAS}
\DeclareMathOperator{\poly}{poly}

\DeclareMathOperator{\psas}{p-SAS}

\DeclareMathOperator{\REG}{REG}

\DeclareMathOperator{\sas}{SAS}

\DeclareMathOperator{\sasRange}{sasRange}
\DeclareMathOperator{\SatProb}{\textsc{CNF-Sat}}
\DeclareMathOperator{\SETH}{\textsf{SETH}}
\DeclareMathOperator{\size}{\mathsf{size}}
\DeclareMathOperator{\smallO}{o}

\DeclareMathOperator{\states}{\mathsf{states}}
\DeclareMathOperator{\Subseq}{Subseq}
\DeclareMathOperator{\subseq}{\preceq}
\DeclareMathOperator{\ScatFact}{\Subseq}

\newcommand{\gapName}{\mathsf{gap}}
\newcommand{\distinctgcsubseq}[3]{\len{#1}_{#2,#3}}
\newcommand{\parikhK}[2]{\Psi_{#2}(#1)}

\newcommand{\gap}[3]{\mathsf{gap}_{#2}(#1, #3)}

\newcommand{\gaptuple}{gc}
\newcommand{\pSubSeqMatch}{\mathtt{pSubSeqMatch}}

\newcommand{\subseqSet}[2]{Subseq(#1, #2)}
\newcommand{\subsequence}[2]{\mathsf{subseq}_{#2}(#1)}

\newcommand{\pMAS}{\mathtt{pMAS}}

\newcommand{\matchProb}{\textsc{Match}}
\newcommand{\nuniProb}{\textsc{NUni}}

\newcommand{\nequiProb}{\textsc{NEqui}}
\newcommand{\uniProb}{\textsc{Uni}}
\newcommand{\contProb}{\textsc{Con}}
\newcommand{\equiProb}{\textsc{Equi}}

\newcommand{\lowerBoundShort}[1]{L^{-}(#1)}
\newcommand{\upperBoundShort}[1]{L^{+}(#1)}

\DeclareMathOperator{\npclass}{NP}

\newtheorem{thm}{Theorem}
\newtheorem{theorem}[thm]{Theorem}
\newtheorem{definition}[thm]{Definition}
\newtheorem{lemma}[thm]{Lemma}
\newtheorem{cor}[thm]{Corollary}
\newcommand{\xSubseq}[1]{#1$-$\Subseq}
\newcommand{\pSubseq}{\xSubseq{p}}
\def\nth#1{#1$^{\text{th}}$}
\newcommand{\len}[1]{|#1|}
\DeclareMathAlphabet{\mathbbb}{U}{bbold}{m}{n}

\title{Combinatorial Algorithms for Subsequence Matching: \\A Survey}
\author{Maria Kosche, \quad Tore Koß, \quad Florin Manea, \quad Stefan Siemer
	\institute{Göttingen University, Germany}
	\email{\{maria.kosche,tore.koss,florin.manea,stefan.siemer\}@cs.uni-goettingen.de}
}
\def\titlerunning{Combinatorial Algorithms for Subsequence Matching: A Survey}
\def\authorrunning{M. Kosche, T. Koß, F. Manea \& S. Siemer}

\begin{document}
\maketitle

\begin{abstract}
In this paper we provide an overview of a series of recent results regarding algorithms for searching for subsequences in words or for the analysis of the sets of subsequences occurring in a word. %	This is a sentence in the abstract.
%	This is another sentence in the abstract.
%	This is yet another sentence in the abstract.
%	This is the final sentence in the abstract.
\end{abstract}

\section{Introduction}

For a string $w = w_1 w_2 \ldots w_n$, where each $w_i$ is a single symbol from some alphabet $\Sigma$, any string $v = w_{i_1} w_{i_2} \ldots w_{i_k}$ with $k \leq n$ and $1 \leq i_1 < i_2 \leq \ldots < i_{k} \leq n$ is called a \emph{subsequence} (also called sometimes \emph{scattered factor} or \emph{subword}) of $w$ (denoted by $v \subseq w$). 

The concept of subsequences is employed in many different areas of computer science. Subsequences appear in areas of theoretical computer science such as, for instance, in formal languages and logics (e.\,g., where they are used in relation to piecewise testable languages~\cite{simonPhD,Simon72,KarandikarKS15,CSLKarandikarS,journals/lmcs/KarandikarS19}, or to define the subword order and downward closures~\cite{HalfonSZ17,KuskeZ19,Kuske20,Zetzsche16}) or in combinatorics on words, where they are used to define the notions of binomial equivalence and binomial complexity, or to introduce the notion of subword histories,   ~\cite{RigoS15,FreydenbergerGK15,LeroyRS17a,Rigo19,Seki12,Mat04,Salomaa05}; however, subsequences are also used in more applied settings, e.\,g., for modelling concurrency~\cite{Riddle1979a, Shaw1978, BussSoltys2014}, or in database theory (especially \emph{event stream processing}~\cite{ArtikisEtAl2017,GiatrakosEtAl2020,ZhangEtAl2014}). Moreover, many classical algorithmic problems are based on subsequences, e.\,g., {longest common subsequence} \cite{DBLP:journals/tcs/Baeza-Yates91} or {shortest common supersequence} \cite{Maier:1978}, and, in particular, such problems have recently regained interest in the context of fine-grained complexity (see~\cite{DBLP:conf/fsttcs/BringmannC18,BringmannK18,AbboudEtAl2015,AbboudEtAl2014}).

There are two main types of algorithmic problems for subsequences investigated in the literature. Firstly, the class of \emph{matching} problems, where one has to decide whether a string $v$ is a subsequence of a string $w$, i.\,e., whether $v \subseq w$ (the name matching comes from the fact that the string $v$ can be seen as a pattern that has to be identified, or matched, within the string $w$). Secondly, the class of \emph{analysis problems}, which are concerned with the algorithmic analysis of the sets $\Subseq_k(w)$ of all length-$k$ subsequences of a given string $w$. Some more concrete examples of analysis problems are the following: for given string $w \in \Sigma^+$ and integer $k \in \mathbb{N}$, we want to decide whether $\Subseq_k(w) = \Sigma^k$ (the \emph{universality} problem), or, for an additional string $v$, we want to decide whether $\Subseq_k(v) = \Subseq_k(w)$ (the \emph{equivalence} problem). For classical subsequences (as defined above), the matching problem is trivial, while the analysis problems are well-investigated and relatively well-understood. In particular, the equivalence problem was introduced by Imre Simon in his PhD thesis \cite{simonPhD}, and was intensely studied in the combinatorial pattern matching community (see \cite{TCS::Hebrard1991,garelCPM,SimonWords,DBLP:conf/wia/Tronicek02,CrochemoreMT03,KufMFCS} and the references therein), before being optimally solved in 2021 \cite{stacs21}. The work on these problems was extended to classes of constrained subsequences, for which very different results were obtained, by fundamentally different methods \cite{Day2022,KKMP22}. \looseness=-1\par

In this work we overview a series of algorithmic, combinatorial, and complexity theoretic results concerning subsequences. For their  original presentation, please see \cite{Barker2020,DayFKKMS21,KufMFCS,stacs21,Kosche2021,KKMP22,Day2022} and the references therein.

Going a bit more into detail, the results surveyed here cover different settings and algorithmic frameworks regarding subsequences, ranging from the classical (and well motivated) case of unrestricted subsequences to some novel ones, where constrained subsequences appear. 

For the case of classical subsequences, the accent of our presentation is put on analysis problems (as matching is trivial). We survey results related to the equivalence and universality problems originally presented in \cite{Barker2020,DayFKKMS21,KufMFCS,stacs21,Kosche2021}. 

For the case of constrained subsequences, we feel that some more discussion about their motivation and origin would be in order. So, the notion of constrained subsequences is rooted in the following main idea: it seems unrealistic for particular scenarios to consider occurrences of $v$ in $w$ where the positions of $w$ that are matching, respectively, the first and last symbol of $v$ (or, similarly, the positions of $w$ matching consecutive symbols of $v$) are very far away from each other. It seems indeed questionable, for instance, whether considering an alignment of DNA-sequences $v$ and $w$ where the nucleotides of $v$ are spread over a factor of $w$ which is several times longer than $v$ itself (or, alternatively, where the nucleotides of $v$ occur in $w$ with arbitrarily long gaps between them) is still meaningful. Similarly, but in a totally different context, when observing a computation, which is represented by a string, one might be more interested in its recent history (and the sequences of events occurring there), rather than analysing the entire computation. Or, in the same setting, one might be interested in sequences of events which occur in a computation, such that the computation executed between two events in such a sequence is constrained by some precise rules. 

In a similar situation, occurring this time in the context of complex event processing \cite{ArtikisEtAl2017,GiatrakosEtAl2020,ZhangEtAl2014}, it might be desirable to describe the situation that between the events of a job $A$, only events associated to a job $B$ appear (e.\,g., due to unknown side-effects this leads to a failure of job $A$). In this case, we are interested in occurrences of a string $v$ as a subsequence of a string $w$ such that the gaps between the positions of $w$ which correspond to the symbols of $v$, only contain symbols from a certain subset of the alphabet (i.\,e., the events associated to job $B$). Moreover, in~\cite{Markus2022}, the authors introduce a query class for event streams, which is essentially based on subsequences with constraints in the form of upper and lower bounds on the length of the gap occurring between consecutive symbols (i.e., events) in the occurrences of the string (i.e., job) $v$ in the larger string/stream $w$. 

Moreover, the fact that in many practical scenarios (including those mentioned above) one has to process streams, which, at any moment, can only be partly accessed by our algorithms, enforces even more the idea that the case where one is interested in subsequences occurring arbitrarily in a given stream (or long string) is less realistic and less useful than the case where one is interested in the subsequences occurring in bounded ranges of the respective stream/string (which can be entirely accessed and processed at any moment by our algorithms). So, wrapping this up, in practice, it makes sense to reason both about the length and the actual content of gaps induced by an occurrence of $v$ in $w$, as well as about the length of the factor (or range) of $w$ in which such an occurrence is contained. 

To this end, in our overview, for the case of constrained subsequences, we will present a series of algorithmic results for problems related to the case of subsequences in which, given strings $w$ and $v$, constraints on either the factors of $w$ in which $v$ may occur as subsequence (called bounded range constraints, see \cite{KKMP22}) are imposed, or constraints on the factors of $w$ occurring between two consecutive letters of an occurrence of $v$ in $w$ (called gap constraints, see \cite{Day2022}) are imposed. In this setting, we overview results regarding the matching problem (which is no longer trivial) as well as results on analysis problems. For the original presentation of these results, as well as for a more detailed overview of the motivations for these particular classes of constrained subsequences and related work, we refer to \cite{KKMP22,Day2022}.

This paper is structured as follows.
We first give a series of general definitions and preliminaries related to subsequences in \cref{sec:defs}.
In \cref{sec:matching},
the matching problem is covered and results are presented for all three cases,
the classical subsequences setting,
for subsequences occurring in bounded ranges,
and for subsequences with gap constraints.
\Cref{sec:analysis} covers the analysis problems in respective subsections, 
namely the universality problem in \cref{sec:universality},
the analysis of absent subsequences in \cref{sec:absent-subsequences},
and the equivalence problem in \cref{sec:equivalence}.
We conclude with a section covering a series of related problems
as well as directions for future work.

\section{Basic Definitions}\label{sec:defs}

Let $\mathbb{N}$ be the set of natural numbers, including $0$. For $m, n \in \mathbb{N}$, we define the range (or interval) of natural numbers lower bounded by $m$ and upper bounded by $n$ as $[m:n] = \{m, m+1, \ldots, n\}$. An alphabet $\Sigma$ is a non-empty finite set of symbols (called letters). A {\em string (or word)} is a finite sequence of letters from $\Sigma$, thus an element of the free monoid $\Sigma^*$. Let $\Sigma^+ = \Sigma^* \setminus \{\varepsilon\}$, where $\varepsilon$ is the empty string. The {\em length} of a string $w \in \Sigma^*$ is denoted by $\len{w}$. The \nth{$i$} letter of $w \in \Sigma^*$ is denoted by  $w[i]$, for $i \in [1:\len w]$. For $a\in \Sigma$, let $\len{w}_a = |\{i \in [1:\len w] \mid w[i] = a \}|$; let $\al(w) = \{x \in \Sigma \mid \len w_x > 0 \}$ be the smallest subset $S \subseteq \Sigma$ such that $w \in S^\ast$. For $m, n \in \mathbb{N}$, with $m\leq n$, we define the range (or factor) of $w$ between positions $m$ and $n$ as $w[m:n] = w[m] w[m+1] \ldots w[n]$. %; clearly, the length of the range $w[m:n]$ is $n-m+1$. 
If $m>n$, then $w[m:n]$ is the empty word. Also, by convention, if $m<1$, then $w[m:n]=w[1:n]$, and if $n>|w|$, then $w[m:n]=w[m:|w|]$. A factor $u=w[m:n]$ of $w$ is called a  {\em prefix} (respectively, {\em suffix}) of $w$ if $m=1$ (respectively, $n=|w|$). \looseness=-1

The powers of a word $w$ are defined as: $w^0=\varepsilon$ and $w^{k+1}=ww^k$, for $k\geq 0$. Define $w^\omega$ as the right infinite word which has $w^n$ as prefix for all $n\geq 0$. The positive integer $p\leq |w|$ is a period of a word $w$ if $w$ is a prefix of $w[1:p]^\omega$. 

We now recall the main notion of this paper, namely the notion of subsequence.
\begin{definition}[Subsequence]
	A word $v$ is a subsequence of length $k$ of $w$ (denoted $v\leq w$), where $\len{w} = n$, if there exist positions $1 \leq i_1 < i_2 < \ldots < i_k \leq n$, such that $v = w[i_1] w[i_2] \cdots w[i_k]$. The set of all subsequences of $w$ is denoted by $\Subseq(w)$. 
\end{definition}

In the following, we will also discuss some other concepts regarding (classical) subsequences,
namely subsequences with gap constraints,
(partitioned into length constraints, regular constraints, or combined length and regular constraints) for the factors of $w$ between two consecutive letters of $v$,
and subsequences within bounded ranges (where we consider, similar to a sliding window scenario,
an integer $p$ an integer $p$ as an upper bound on the range of a word in which a subsequence may occur).

Firstly, we introduce subsequences with gap constraints, or gapped subsequences. This presentation is based on \cite{Day2022}.
We begin by defining the notion of gap constraints. We recall that for a string $w$,
an embedding is a function $e : [k] \to [|w|]$ such that $i < j$ implies $e(i) < e(j)$ for all $i, j \in [k]$,
and it \emph{induces the subsequence $\subsequence{w}{e} = w[e(1)]w[e(2)]\ldots w[e(k)]$ of $w$}. 
For every $j \in [k-1]$, the \emph{$j^{\text{th}}$ gap of $w$ induced by $e$}
is the string $\gap{w}{e}{j} = w[e(j)+1..e(j+1)-1]$.
We say that $e$ is the embedding of $\subsequence{w}{e}$ in $w$.

\begin{definition}[Gap constraints]
	An \emph{$\ell$-tuple of gap constraints} is a tuple $\gaptuple = (C_1, C_2, \ldots, C_{\ell})$
	with $C_i \subseteq \Sigma^*$ for every $i \in [\ell]$.
	For convenience, we set $\gaptuple[i] = C_i$ for every $i \in [\ell]$.
	We say that an embedding $e$ \emph{satisfies a $(k-1)$-tuple of gap constraints $\gaptuple$ with respect to a string $w$}
	if it has the form $e: [k] \to [|w|]$,
	and, for every $i \in [k - 1]$, $\gap{w}{e}{i} \in C_i$.
	For a $(k-1)$-tuple $\gaptuple$ of gap constraints,
	the set $\subseqSet{\gaptuple}{w}$ contains all subsequences of $w$ induced by embeddings that satisfy $\gaptuple$,
	i.\,e., $\subseqSet{\gaptuple}{w} = \{\subsequence{w}{e} \mid e \text{ is an embedding that satisfies } \gaptuple \text{ w.\,r.\,t. $w$}\}$.
	The elements of $\subseqSet{\gaptuple}{w}$ are also called the \emph{$\gaptuple$-subsequences of $w$}.
	
	For a $(|u|-1)$-tuple $\gaptuple$ of gap constraints,
	we write $u \subseq_{\gaptuple} v$ to denote that $u \subseq_e v$
	for some embedding $e : [|u|] \to [|v|]$
	that satisfies $\gaptuple$ with respect to $v$,
	i.\,e., $u \subseq_{\gaptuple} v$ means that $u$ is a $\gaptuple$-subsequence of $v$.
\end{definition}

We generally distinguish the following types of gap constraints:
\begin{itemize}
	\item \emph{regular constraints} if $C_i \in \REG$ for every $i \in [k-1]$. For every $i \in [k-1]$, the regular constraint $C_i$ is represented by a deterministic finite automaton (for short, $\DFA$) $A_i$ accepting it. 
	\item \emph{length constraints} if, for every $i \in [k-1]$, there are $\lowerBoundShort{i}, \upperBoundShort{i} \in \mathbb{N}\cup\{0,+\infty\}$ with $\lowerBoundShort{i} \leq \upperBoundShort{i}$, such that $C_i = \{v \in \Sigma^* \mid  \lowerBoundShort{i} \leq |v| \leq \upperBoundShort{i}\}$. Length constraints are succinctly represented by pairs of numbers $(\lowerBoundShort{i}, \upperBoundShort{i})$, $i \in [k-1]$, in binary encoding. 
	\item \emph{reg-len constraints} if, for every $i \in [k-1]$, $C_i$ is the conjunction of a regular constraint $C'_i$ and a length constraint $(\lowerBoundShort{i}, \upperBoundShort{i})$, i.\,e., $C_i = C'_i \cap \{v \in \Sigma^* \mid  \lowerBoundShort{i} \leq |v| \leq \upperBoundShort{i}\}$. Such constraints are represented by $((\lowerBoundShort{i}, \upperBoundShort{i}),A'_i)$, where $A'_i$ is a $\DFA$ accepting $C'_i$.
\end{itemize}

We move now further, and introduce the concept of $p$-subsequence, or subsequences occurring within bounded ranges. For this presentation, we follow \cite{KKMP22}.
 
\begin{definition}[Bounded range constraints]
	1. Let $v,w\in\Sigma^*$ with $\len v=m,\len w=n$. The string $v$ is called a $p$-subsequence of $w$ (denoted $v \leq_p w$) if there exists an embedding $e:[m]\to [n]$ such that $v=\subsequence{w}{e}$ and $\len{w[e(1):e(m)]}\leq p$, or equivalently $e(m)-e(1)\leq p-1$.
%$i \leq \len{w} - p + 1$ such that $v$ is a subsequence of $w[i:i + p - 1]$.\\
	2. For $p\in\mathbb N$ and $w\in\Sigma^*$, we denote the set of all $p$-subsequences of $w$ by $\pSubseq(w) = \{v\in \Sigma^*\mid v \leq_p w\}$. Furthermore, for $k\in \mathbb N$, we denote the set of all $p$-subsequences of length $k$ of $w$ by $\pSubseq_{k}(w)$.
\end{definition}

Once these main concepts introduced, we can now discuss several preliminaries which are necessary for understanding the surveyed results.

\smallskip

\noindent\emph{Computational Model.} In general, the problems surveyed here are of algorithmic nature. The computational model used to describe the algorithms is the standard unit-cost RAM with logarithmic word size: for an input of size $N$, each memory word can hold $\log N$ bits. Arithmetic and bitwise operations with numbers in $[1:N]$ are, thus, assumed to take $\bigO(1)$ time. %Numbers larger than $N$, with $\ell$ bits, are represented in $\bigO(\ell/\log N)$ memory words, and working with them takes time proportional to the number of memory words on which they are represented. 
In all the problems, it is assumed that we are given a word $w$ or two words $w$ and $u$, with $|w|=n$ and $|v|=m$ (so the size of the input is $N=n+m$), over an alphabet $\Sigma=\{1,2,\ldots,\sigma\}$, with $2\leq |\Sigma|=\sigma\leq n+m$. That is, the processed words are assumed to be sequences of integers (called letters or symbols), each fitting in $\bigO(1)$ memory words. This is a common assumption in string algorithms: the input alphabet is said to be {\em an integer alphabet}. For more details see, e.\,g.,~\cite{crochemore}.

The algorithmic results (upper bounds) that are surveyed here are complemented by a series of lower bounds. In those cases, the results hold already for the case of constant alphabets. That is, they hold already when the input of the problem is restricted to words over an alphabet $\Sigma=\{1,2,\ldots,\sigma\}$, with $\sigma\in \bigO(1)$. 

\smallskip

\noindent\emph{Complexity Hypotheses.} 
For the series of conditional lower bounds for the time of complexity of the considered problems, we now recall some standard computational problems and complexity hypotheses regarding them, respectively, on which the proofs of lower bounds are based.

The \emph{Satisfiability problem for formulas in conjunctive normal form}, $\SatProb$, gets as input a Boolean formula $F$ in conjunctive normal form as a set of clauses $F = \{c_1, c_2, \ldots, c_m\}$ over a set of variables $V = \{v_1, v_2, \ldots, v_n\}$, i.\,e., for every $i \in [m]$, we have $c_i \subseteq \{v_1, \neg v_1, \ldots, v_n, \neg v_n\}$. The question is whether $F$ is satisfiable. 
%, i.\,e., whether there is an assignment $\pi : V \to \{0, 1\}$ that makes at least one literal of each clause $c_i$ true. 
By $k$-$\SatProb$, we denote the variant where $|c_i| \leq k$ for all $i \in [m]$.\looseness=-1

The \emph{Orthogonal Vectors problem} ($\OV$ for short) gets as input two sets $A, B$ each containing $n$ Boolean-vectors of dimension~$d$, where $d\in \omega(\log n)$. The question is whether there exist two vectors $\vec{a} \in A$ and $\vec{b} \in B$ which are orthogonal, i.\,e., $\vec{a}[i] \cdot \vec{b}[i] = 0$ for every $i \in [d]$.

We shall use the following algorithmic hypotheses based on $\SatProb$ and $\OV$
that are common for obtaining conditional lower bounds in fine-grained complexity.
In the following, $\poly$ is any fixed polynomial function: 

\smallskip
\noindent -- \emph{Exponential Time Hypothesis}  ($\ETH$)~\cite{ImpagliazzoEtAl2001,LokshtanovEtAl2011}: $3$-$\SatProb$ cannot be solved in time $2^{\smallO(n)} \poly(n + m)$.
%, where 
%$n$ is the number of variables and 
%$\poly$ is any fixed polynomial function.

\noindent -- \emph{Strong Exponential Time Hypothesis} ($\SETH$)~\cite{ImpagliazzoPaturi2001,Williams2015}: For every $\epsilon > 0$ there exists $k$ such that $k$-$\SatProb$ cannot be decided in $\bigO(2^{n(1-\epsilon)} \poly(n))$ time. %, where 
%$n$ is the number of variables and 
%$\poly$ is any fixed polynomial function.

The following result, which essentially formulates the Orthogonal Vectors Hypothesis ($\OVH$), can be shown (see~\cite{Bringmann2014,Bringmann2019,Williams2015}).
\begin{lemma}
	$\OV$ cannot be solved in $\bigO(n^{2-\epsilon} poly(d))$ time for any $\epsilon > 0$, unless $\SETH$ fails. 
\end{lemma}

\section{Matching problems}\label{sec:matching}

The matching problem $\matchProb$ for subsequences is to decide,
given two words $u,w\in\Sigma^*$ with $\len u=m$ and $\len w=n$,
whether $u$ is a subsequence of $w$.
In the general case,
that is without further restrictions like gap constrains or bounded ranges,
it is quite easy to answer:
if we go left-to-right through $w$ and greedily search for the letters $u[1]$ to $u[m]$,
we answer positively if and only if we find all letters from $u$ in $w$.
This greedy approach obviously is correct and works in linear time $\bigO(n)$. 

\paragraph{Considering bounded range constraints.} If we consider subsequences occurring within bounded ranges, the problem changes as follows: for $u,w$ as above and $p\in\mathbb N$ with $p\geq m$, we need to decide whether $u$ is a $p$-subsequence of $w$. Simply using the greedy approach above for each range still works, but is not optimal anymore (it has $\bigO(np)$ complexity). However, by reading the word $w$ left to right and maintaining an array which saves for every $i\in [1:m]$ the length of the shortest suffix of the current range $w[t-p+1:t]$ containing $u[1:i]$ (if there is any) and updating the array when we increment $t$ (i.\,e., read the a new letter of the word $w$), we can reduce the time complexity to $\bigO(mn)$ (see \cite{KKMP22} for a more detailed description of the algorithm).

\begin{theorem}\label{thm:algoMatching}
	$\matchProb$ in bounded ranges can be solved in $\bigO(mn)$ time.
\end{theorem}

The algorithm presented in \cite{KKMP22} can be, in fact, seen as an algorithm in the sliding window model with window of fixed size $p$ (see \cite{GanardiHL16,GanardiHKLM18,GanardiHLS19}). More precisely, it scans the stream $w$ left to right and, when the $t^{th}$ letter of the stream is scanned, it reports whether the window $w[t-p+1:t]$ contains $u$ as a subsequence. In other words, it reports whether the string $w[t-p+1:t]$ is in the regular language $L_u=\{v\mid u\leq v\}$. The problem of checking whether the factors of a stream scanned by a sliding window are in a regular language was heavily investigated, see \cite{Ganardi19} and the references therein. In particular, from the results of \cite{GanardiHL16} it follows that, for a constant $u$ (i.\,e., $u$ is not part of the input), the problem of checking whether the factors of a stream scanned by a sliding window are in the language $L_u$ cannot be solved using $\smallO(\log p)$ bits when the window size is not changing and equals $p$. We note that the algorithm of \cite{KKMP22} is optimal from this point of view: if $u$ is constant and, thus, $m\in \bigO(1)$, it uses $\bigO(\log p)$ bits to store the maintained data structures. 

Moreover, the algorithm presented in \cite{KKMP22} is optimal also from the time complexity point of view, unless $\OVH$ fails.

\begin{theorem}\label{thm:lowerBoundSubseq}
	$\matchProb$ in bounded ranges cannot be solved in time $\bigO(n^h m^g)$, where $h+g= 2-\epsilon$ with $\epsilon>0$, conditional to $\OVH$. 
\end{theorem}

\paragraph{Considering gap constraints.} For subsequences with gap constraints, the matching problem is to decide, for given strings $u$, $w$, and gap constraints $\gaptuple$ with $|\gaptuple|=|u|-1$, whether $u$ is a $\gaptuple$-subsequence of $w$ (i.\,e., whether $u \in \subseqSet{\gaptuple}{w}$).

The results from \cite{Day2022} give us the following upper bound in the case of reg-len constraints
where $\gaptuple$ denotes the given gap constraints,
$\states(\gaptuple)$ denotes the total number of states of the $\DFA$s that represent the regular constraints,
$\size(\gaptuple)$ is the total size of the automata defining these constraints and $\nz(\gaptuple)$ is the number of gaps which are not equal to $\{\epsilon\}$.

\begin{theorem}\label{constantPatternsRegularLength}
	$\matchProb$ with reg-len constraints can be solved in $\bigO(|w|\states(\gaptuple) + \size(\gaptuple))$~time.\looseness=-1
\end{theorem}

The proof given in \cite{Day2022} is based on a dynamic programming approach, which is implemented in the respective time complexity with the help of some relatively involved data structures, and implies that when considering only length constraints or only regular constraints (and not combined reg-len constraints), the following rectangular upper bounds hold.

\begin{cor}\label{constantPatternsRL}
	(1). $\matchProb$ with length constraints can be solved in $\bigO(|w|\nz(\gaptuple))$ time.\\ (2). $\matchProb$ with regular constraints can be solved in $\bigO(|w|\states(\gaptuple) + \size(\gaptuple))$ time.
\end{cor}

When $\gaptuple$ only consists of constraints that are $\{\varepsilon\}$ or $\Sigma^*$, respectively,
the case of string matching or, respectively, subsequence matching is modelled,
which can be solved in linear time. As far as length constraints are concerned,
it seems that non-trivial upper bounds lead to an increase in the difficulty of the $\matchProb$ problem;
a particularly efficient approach for subsequences with general length constraints is given in \cite{BilleEtAl2012}
but in the worst case it still has
rectangular complexity.
However, even when non-trivial length upper bounds are used,
there are still some simpler particular cases.
For instance, when working with strings with \emph{don't cares}
(or partial words), where each gap has a fixed length
(i.\,e., the lower and upper bounds are the same),
Match can be solved in time $\bigO(|w| \log |p|)$ \cite{CliffordC07}.

A reduction from the $\OV$ problem is given in \cite{Day2022},
which shows that $\matchProb$ for non-trivial length or regular constraints is more difficult than $\matchProb$ for the classical subsequences scenario.
The following conditional lower bounds
for subsequences with length and/or regular gap constraints are obtained.

\begin{theorem}\label{lowerBoundLength}
	$\matchProb$ with length constraints cannot be solved in $\bigO(|w|^h \nz(\gaptuple)^g)$ time 
	with $h+g= 2-\epsilon$ for some $\epsilon>0$, unless $\OVH$ fails. This holds even if $|\Sigma| = 4$ and all length constraints are $(0, \ell)$ with $\ell \leq 6$.
\end{theorem}

\begin{cor}\label{lowerBoundRegular}
	$\matchProb$ with regular constraints
	cannot be solved in $\bigO(|w|^h \states(\gaptuple_p)^g)$ time with $h+g= 2-\epsilon$ for some $\epsilon>0$, unless $\OVH$ fails. This holds even if $|\Sigma| = 4$ and all regular constraints are expressed by constant size DFAs.
\end{cor}

\section{Analysis problems}\label{sec:analysis}
When considering algorithmic analysis problems related to subsequences, typical research questions are concerned with structural properties of the set of all (constrained) subsequences occurring in a word, as well as with finding minimal (w.\,r.\,t. length or w.\,r.\,t. the subsequence relation) missing subsequences of a word.

\subsection{Universality}\label{sec:universality}

Generally speaking,
the universality problem $\uniProb$ is to decide,
for given integer $k$ and string $w \in \Sigma^*$ with $\len w = n$,
whether the set of subsequences of length $k$ of $w$ equals the set $\Sigma^k$. For convenience, we will also consider in the following the complement problem, i.\,e., \emph{non-universality problem} ($\nuniProb$). 
%
%In the following,
%we first define some notions which are important for the discussion of this section.

\begin{definition}
	A word $w\in\Sigma^{\ast}$ is called {\em $k$-subsequence universal}  (w.\,r.\,t. $\Sigma$, for short {\em $k$-universal}), for $k\in\N$, if $\ScatFact_k(w)=\Sigma^k$. We abbreviate $1$-universal by {\em universal}. The {\em universality-index} $\iota(w)$ of $w\in\Sigma^{\ast}$ is the largest $k$ such that $w$ is $k$-universal.
\end{definition}

If $\iota(w)=k$ then $w$ is $\ell$-universal for all $\ell\leq k$. Notice that $k$-universality is always w.\,r.\,t. a given alphabet $\Sigma$: the word $\ta\tb\tc\tb\ta$ is universal for $\Sigma=\{\ta,\tb,\tc\}$ but it is not universal for $\Sigma\cup\{\td\}$. 

The notion of $k$-universality coincides to that of $k$-richness introduced in \cite{CSLKarandikarS,journals/lmcs/KarandikarS19}. We use the name {\em $k$-universality} rather than {\em $k$-richness}, as richness of words is also used with other meanings, see, e.\,g., \cite{DroubayJP01,LucaGZ08}. 
We recall the arch factorisation, introduced by Hebrard~\cite{TCS::Hebrard1991}.

\begin{definition}[\cite{TCS::Hebrard1991}]\label{archfact}
	For $w\in\Sigma^{\ast}$ the {\em arch factorisation} of $w$ is $w=\ar_w(1)\cdots \ar_w(k)r(w)$ for some $k\in\N_0$ where $\ar_w(i)$ is universal,  
	the last letter of $\ar_{w}(i)$, namely $\ar_w(i)[|\ar_w(i)|]$, does not occur in $\ar_w(i)[1:|\ar_w(i)|-1]$  for all $i\in[1:k]$, and 
	$\letters(r(w))\subset\Sigma$. 
	The words $\ar_w(i)$ are called {\em arches} of $w$, $r(w)$ is called the {\em rest}. % Set $m(w)=\ar_w(1)[|\ar_w(1)|] \cdots \ar_w(k)[|\ar_w(k)|]$ as the word containing the unique last letters of each arch.
\end{definition}

If the arch factorisation of $w$ contains $k\in\N_0$ arches, then $\iota(w)=k$. The arch factorization of a word $w$ can be computed in linear time and, as such, we could check in linear time if a given word $w$ is $k$-universal (see, e.\,g., \cite{Barker2020}). 
The following immediate theorem based on the work of Simon \cite{Simon72} completely characterises the set of $k$-subsequence universal words, based on Hebrard's arch factorisation.

\begin{thm}\label{theodlt}
	The word  $w\in\Sigma^{\ast}$ is $k$-universal if and only if there exist the words $v_i$, with $i\in [1:k]$, such that $v_1\cdots v_k=w$ and $\letters(v_i)=\Sigma$ for all $i\in [1:k]$.
\end{thm}

This property gives us some insight in the combinatorial structure of the subsequences occurring inside words. For instance, it can be used to directly compare two words w.\,r.\,t. their universality index, or serve as a starting point for the analysis of the set of missing subsequences of words. See, for instance, \cite{Barker2020,DayFKKMS21,Fleischmann2022,FleischmannDCFS2022,Day2022,KKMP22}.

\paragraph{The Edit Distance to \texorpdfstring{$k$}{k}-Subsequence Universality.}
As a natural extension to the universality property from above, as one can do for almost every string property, we can ask how far is a word from fulfilling that property. That is, we can ask for the distance from a give string to the set of strings which fulfill that property,  with respect to some string metric. More precisely in this section, we discuss how to compute the minimal number of edits we need to apply to a word $w$, with $|w|=n$, $\letters(w)=\Sigma$, with universality index $\iota(w)$, so that it is transformed into a word with universality index $k$, w.\,r.\,t. the same alphabet $\Sigma$. The edits considered are insertion, deletion, substitution, and the number we want to compute can be seen as the {\em edit distance} between $w$ and the set of $k$-universal words over $\Sigma$. 

The first thing that we can see is that, if we want to obtain a $k$-universal word with $k>\iota(w)$, then it is enough to consider only insertions. Indeed, deleting a letter of a word can only restrict the set of subsequences of the respective word, while in this case we are interested in enriching it. Substituting a letter might make sense, but it can be simulated by an insertion: assume one wants to substitute the letter $\ta$ on position $i$ of a word $w$ by a $\tb$. It is enough to insert a $\tb$ next to position $i$, and the set of subsequences of $w$ is enriched with all the words that could have appeared as subsequences of the word where $\ta$ was actually replaced by $\tb$. We might, in the end, have some extra words in the set of subsequences, which would have been eliminated through the substitution, but it does not affect our goal of reaching $k$-universality. 

If we want to obtain a word with universality index $k$, for $k<\iota(w)$, then it is enough to consider only deletions. Assume that we have a sequence of edits that transforms the word $w$ into a word $w'$ with universality index $k$. Now, remove all the insertions of letters from that sequence. The word $w''$ we obtain by executing this new sequence of operations clearly fulfils $\iota(w'')\leq \iota(w')$. Further, in the new sequence, replace all substitutions with deletions. We obtain a word $w'''$ with a set of subsequences strictly included in the one of $w''$, so with $\iota(w''')\leq \iota(w'')$. As each deletion changes the universality index by at most $1$, it is clear that (a prefix of) this new sequence of deletions witnesses a shorter sequence of edits which transforms $w$ into a word of universality index $k$. 

So, to increase the universality index of a word it is enough to use insertions and to decrease the universality index of a word it is enough to use deletions.
Nevertheless, one might be interested in what happens if we only 
use substitutions. In this way, we can both decrease and increase the 
universality index of a word. Moreover, one can see the minimal number of 
substitutions needed to transform $w$ into a $k$-universal word as the {Hamming distance} between $w$ and the set of $k$-universal words.
In the following we list all of the resulting theorems individually. All of the results are achieved by a dynamic programming approach combined with a sophisticated analysis of the combinatorial properties of $k$-universal words and some new specialized data structures. See the full proofs and algorithms in \cite{DayFKKMS21}.

\begin{thm}\label{ins-distance}
	Let $w$ be a word, with $|w|=n$, $\letters(w)=\Sigma$, and $\Sigma = \{1,2,\ldots,\sigma\}$. Let $k\geq \iota(w)$ be an integer. We can compute the minimal number of insertions needed to apply to $w$ in order to obtain a $k$-universal word (w.\,r.\,t. $\Sigma$) in $\bigO(nk)$~time if $k\leq n$ and $\bigO(T(n,\sigma,k))$ time otherwise, where $T(n,\sigma,k)$ is the time needed to compute the number $k\sigma-n $. 
\end{thm}

\begin{thm} \label{deletions}
	Let $w$ be a word, with $|w|=n$, $\letters(w)=\Sigma$, and $\Sigma = \{1,2,\ldots,\sigma\}$. 
	Let $k$ be an integer with $k \leq \iota(w)\leq \lfloor \frac{n}{\sigma} \rfloor $. We can compute in $\bigO(nk)$ time the minimal number of deletions needed to obtain a word of universality index~$k$ (w.\,r.\,t. $\Sigma$) from $w$.
\end{thm}

\begin{thm}\label{edit-subs}
	Let $w$ be a word, with $|w| = n$, $alph(w)=\Sigma $, and $\Sigma= \lbrace 1, 2, \ldots, \sigma \rbrace$. Let $k$ be an integer $0\leq k \leq \lfloor \frac{n}{\sigma} \rfloor $. We can compute the minimal number of substitutions needed to apply to $w$ in order to obtain a $k$-universal word (w.\,r.\,t. $\Sigma$) in $\bigO(nk)$ time.
\end{thm}

\paragraph{Universality and bounded range constraints.}

After looking into the (unrestricted) subsequence universality of a word and in particular the edit distance between a word and the set of $k$-universal words, we will now discuss the case of subsequences occurring in bounded ranges or words. In this case, the universality problem $\uniProb$ asks to decide for given word $w$, alphabet $\Sigma$, and integers $k$ and $p$, with $\len w = n$ and $k \leq p \leq n$, whether $\pSubseq_{k}(w) \neq \Sigma^k$.

Surprisingly, in this case, we get an intractability result, complemented by a fine-grained lower bound (see \cite{KKMP22}). For convenience, this computational hardness result is stated for $\nuniProb$. 

\begin{theorem}\label{thm:univ-np-hard}
	$\nuniProb$ for bounded ranges is NP-hard and cannot be solved in subexponential time $2^{\smallO(k)} \poly(k,n)$ unless $\ETH$ fails.
\end{theorem}

The proof for this result involves the reduction from the related problem $\nuniProb$ for partial words,
which asks to decide, for given list of partial words $S = \{w_1, \ldots, w_k\}$ over $\{0,1\}$, where every partial word has same length $L$,
whether there exists a word $v \in \{0,1\}^L$ such that $v$ is not compatible with any of the partial words in $S$
(while in this context, two partial words $u$ and $v$ of the same length are \emph{compatible} if, for all $i\in [|u|]$, we have that either $u[i]=v[i]$ or at least one of $u[i]$ or $v[i]$ is undefined).

\paragraph{Universality and gap constraints.}
For the case of subsequences with gap constraints,
\cite{Day2022} presents a series of results starting with a brute force upper bound
which can be derived from the results for $\matchProb$ for subsequences with gap constraints.

\begin{theorem}\label{upperBoundUnboundedAlphabets}
	(1) The problem $\uniProb$
	%$, \contProb$ and $\equiProb$ 
	for subsequences with length (or reg-len) constraints can be solved in time $\bigO(|\Sigma|^{k} \nz(\gaptuple) \ell)$
	(respectively, $\bigO(|\Sigma|^{k} \states(\gaptuple) \ell)$),
	where $\ell = \max\{|w|, |w'|\}$.\\
	(2) For the case of a fixed alphabet $\Sigma$ (i.\,e., $|\Sigma|\in \bigO(1)$), the problem $\uniProb_{\Sigma}$
	%, $\contProb_{\Sigma}$ and $\equiProb_{\Sigma}$
	with length (or reg-len) constraints can be solved in time $2^{\bigO(k)} \nz(\gaptuple) \ell$ (respectively, $2^{\bigO(k)} \states(\gaptuple) \ell$), where $\ell = \max\{|w|, |w'|\})$.
\end{theorem}

At the same time,
the following lower bound (again, given for $\nuniProb$) shows that it is unlikely for significantly faster algorithms to exist.

\begin{theorem}\label{HardnessNonUniversalityBounded}
	For every fixed alphabet $\Sigma$ with $|\Sigma| \geq 3$, $\nuniProb_{\Sigma}$
	%, $\ncontProb_{\Sigma}$ and $\nequiProb_{\Sigma}$ 
	with length constraints is $\npclass$-complete,
	even if all length constraints are $(1, 5)$. Moreover,
	\begin{itemize}
		\item it cannot be solved in subexponential time $2^{\smallO(k)} \poly(|w|, k))$ (unless ETH fails), 
		\item it cannot be solved in time $\bigO(2^{k(1-\epsilon)} \poly(|w|, k))$ (unless SETH fails).
	\end{itemize}

	For a fixed alphabet $\Sigma$ with $|\Sigma| = 2$, $\nuniProb_{\Sigma}$
	%, $\ncontProb_{\Sigma}$ and $\nequiProb_{\Sigma}$
	with length constraints is $\npclass$-complete even if each length constraint is $(0, 0)$ or $(3, 9)$
	(meaning that each gap is either empty
	or has length between $3$ and $9$).\looseness=-1
\end{theorem}
%\begin{theorem}\label{HardnessNonUniversalityUnbounded}
%	Problems $\nuniProb$, $\ncontProb$ and $\nequiProb$ with length constraints cannot be solved in running time $\bigO(f(k) \poly(|w|, k))$ for any computable function $f$ (unless $\fptclass = \wclass[1]$). 
%\end{theorem}

\subsection{Absent Subsequences}\label{sec:absent-subsequences}

In the previous sections we surveyed a series of results related to deciding whether a string contains as subsequences all strings of length up to $k$. Now, we focus on understanding the strings which do not occur as subsequences of a given input string.

So, in this subsection we summarize a series of algorithmic and complexity results related to decision problems concerning shortest and, respectively, minimal absent subsequences. Once more, we begin with the classical case, and then discuss the case of subsequences occurring within bounded ranges.

We begin with several definitions. 

\begin{definition}[-- Absent subsequences] 
	A word $v$ is an absent subsequence of $w$ if $v$ is not a subsequence of $w$.
	An absent subsequence $v$ of $w$ is a minimal absent subsequence (for short, $\mas$) of $w$ if every proper subsequence of $v$ is a subsequence of $w$.
	We will denote the set of all $\mas$ of $w$ by $\mas(w)$.
	An absent subsequence $v$ of $w$ is a shortest absent subsequence (for short, $\sas$) of $w$ if $|v|\le |v'|$ for any other absent subsequence $v'$ of $w$.
	We will denote the set of all $\sas$ of $w$ by $\sas(w)$.
\end{definition}

\paragraph{Absent subsequences in words.}

Note that, in general, any shortest absent subsequence of a word $w$ has length $\iota(w) + 1$, where $\iota(w)$ is the universality index of $w$. This already establishes a connection to the results presented in previous section. Moreover, we notice that we can easily find at least one $\sas$ (and therefore an $\mas$) of $w$ from its arch factorisation: for $1\leq i\leq \iota(w)$ let $u_i = \ar_w(i)[\len{\ar_w(i)]}$ be the last letter of the \nth i arch of $w$ and $u_{\iota(w)+1}$ be any letter not occurring in $r(w)$, then $u=u_1\cdots u_{\iota(w)+1}$ is an $\sas$ of $w$. By refining this approach we can, for given $w\in\Sigma^*$, build in linear time a data structure allowing us to identify a succinct representation of a $\sas$ of any factor $w[i:j]$ of $w$ in constant time (and effectively output this $\sas$ in time proportional to its length).  

\begin{theorem}[\cite{Kosche2021}]
	For a word $w$ of length $n$ we can construct in $\bigO(n)$ time data structures allowing us to answer in $\bigO(1)$ time  queries {\em $\sasRange(i,j)$: ``return a representation of an $\sas$ of $w[i:j]$". }
\end{theorem}

Given $u,w\in\Sigma^*$ the problem to check whether $u$ is an $\sas$ or, respectively, an $\mas$ of $w$ is decidable in linear time (see \cite{Kosche2021}).

\begin{theorem}
	Given a word $w$ of length $n$ 
	and a word $u$ of length $m$,
	we can test in $\bigO(n)$ time whether $u$ is an $\sas$ or $\mas$ of $w$. 
\end{theorem}

In both cases, we check trivially whether $u$ is absent from $w$ or not. In the case we want to decide whether $u$ is an $\sas$ of $w$ we simply check that $\len u=\iota(w)+1$. When checking whether $u$ is an $\mas$ we calculate the shortest prefixes of $w$ containing $u[1:1], u[1:2],\ldots,u[1:m-1]$ respectively, as well as the shortest suffixes of $w$ containing $u[m:m], u[m-1:m],\ldots, u[2:m]$ respectively. Now, $u$ is an $\mas$ of $w$ if and only if for every $1\leq i\leq m$ the shortest prefix containing $u[1:i-1]$ and the shortest suffix containing $u[i+1:m]$ do not overlap.

For the analysis of the set $\sas(w)$ we can construct in linear time data structures, visualized by a tree called $\sas$-tree, which encodes $\sas(w)$.

\begin{theorem}[\cite{Kosche2021}]
	Given a word $w$ of length $n$ with universality index $k$, we can construct in $\bigO(n)$ time data structures allowing us to perform the following tasks:
	\begin{enumerate}
		\item We can check in $\bigO(k)$ time if a word $u$ of length $k+1$ is an $\sas$ of $w$. 
		\item We can compute in $\bigO(k)$ time the lexicographically smallest $\sas$ of $w$.
		\item We can efficiently enumerate (i.\,e., with polynomial delay) all the $\sas$ of $w$.
	\end{enumerate}
\end{theorem}

For minimal absent subsequences the problem becomes more complicated but we can still construct data structures encoding $\mas(w)$, visualized by a directed acyclic graph called $\mas$-DAG, in $\bigO(n^2\sigma)$ time.
It is worth noting here that the lexicographic smallest $\mas$ of $w$ is $a^{\len w_a+1}$ where $a$ is the lexicographic smallest letter of $\Sigma$ which occurs in $w$.
\begin{theorem}[\cite{Kosche2021}]
	For a word $w$, we can construct in $\bigO(n^2\sigma)$ time data structures allowing us to efficiently perform the following tasks: 
	\begin{enumerate}
		\item We can check in $\bigO(m)$ time if a word $u$ of length $m$ is an $\mas$ of $w$. 
		\item We can compute in polynomial time the longest $\mas$ of $w$. 
		\item We can check in polynomial time for a given length $\ell$ if there exists an $\mas$ of length $\ell$ of $w$. 
		\item We can efficiently enumerate (with polynomial delay) all the $\mas$ of $w$.
	\end{enumerate}
\end{theorem}

In the end of this subsection we present  a result allowing us to check whether a word $u$ can be extended to an $\mas$, that is checking whether there is an $\mas$ of $w$ having $u$ as a prefix, and if possible calculates the shortest such $\mas$. 

\begin{cor}
	For a word $w$ of length $n$, we can construct in $\bigO(n\sigma)$ time data structures allowing us to answer $\masExtend(u)$ queries: for a subsequence $u$ of $w$, decide whether there exists an $\mas$ $uv$ of $w$, and, if yes, construct such an $\mas$ $uv$ of minimal length. The time needed to answer a query is $\bigO(|v|+|u|)$. 
\end{cor}

\paragraph{Absent subsequences and bounded range constraints.}

We continue by considering the case when bounded range restrictions are added in the study of absent subsequences. That is, we are only interested in subsequences not occurring in any factor of fixed length $p$ of $w$ (but, which may occur in longer factors). We call such a sequence \emph{absent $p$-subsequence}. Similarly we define the notions of \emph{shortest absent $p$-subsequences} $\psas$ and \emph{minimal absent $p$-subsequences} $\pmas$. 

\begin{definition}[-- Absent $p$-subsequence]
	The word $v$ is an absent $p$-subsequence of $w$ if $v \notin \pSubseq(w)$. We also say $v$ is $p$-absent from $w$.
	The word $v$ is a $\psas$ (shortest absent $p$-subsequence) of $w$ if $v$ is an absent $p$-subsequence of $w$ of minimal length.
	The word $v$ is a $\pmas$ (minimal absent $p$-subsequence) of $w$ if $v$ is an absent $p$-subsequence of $w$ but all subsequences of $v$ are $p$-subsequences of $w$.
\end{definition}

Adding the bounded range restriction to absent subsequences complicates significantly some of the algorithmic tasks which were efficiently solved in the original setting. In particular, checking whether $u$ is not a $p$-$\sas$ of $w$ is NP-hard and cannot be computed in subexponential time (conditional to $\ETH$).

\begin{theorem}
	Deciding whether $v$ is not a $p$-$\sas$ of $w$ is NP-hard and cannot be solved in subexponential time $2^{\smallO(k)} \poly(k,n,m)$ unless $\ETH$ fails.
\end{theorem}

We denote by $\pMAS$ the decision problem to check for given strings $v,w\in\Sigma^*$ whether $v$ is a $p$-$\mas$ of $w$. $\pMAS$ is still decidable in polynomial time $\bigO(\len u\len w)$, which is also optimal unless $\OVH$ fails. An optimal algorithm is given in \cite{KKMP22}.  
%instead of the linear time algorithm for the general case.

\begin{theorem}\label{thm:algpmas}
	$\pMAS$ can be solved in time $\bigO(nm)$, where $\len v = m, \len w = n$.
\end{theorem}

\begin{theorem}
	$\pMAS$ cannot be solved in time $\bigO(n^h m^g)$ where $h+g= 2-\epsilon$ with $\epsilon>0$, unless $\OVH$ fails. 
\end{theorem}

Similarly to the case of \cref{thm:algoMatching}, the algorithm proposed in \cite{KKMP22} can be seen as working in the sliding window model, with window of fixed size $p$. If, as in the case of the discussion following \cref{thm:algoMatching}, we assume $u$ (and $m$) to be constant, we obtain a linear time algorithm. However, its space complexity, measured in memory words, is $\bigO(p)$ (as we need to keep track, in this case, of entire content of the window). In fact, when $m$ is constant, it is easy to obtain a linear time algorithm using $\bigO(1)$ memory words (more precisely, $\bigO(\log p)$ bits of space) for this problem: simply try to match $u$ and all its subsequences of length $(m-1)$ in $w$ simultaneously, using the algorithm from \cref{thm:algoMatching}. Clearly, $u$ is a $\pmas$ if and only if $u$ is not a subsequence of $w$, but all its subsequences of length $m-1$ are. However, the constant hidden by the ${\bigO}$-notation in the complexity of this algorithm is proportional with $m^2$. It remains open whether there exists a (sliding window) algorithm for  $\pMAS$ both running in ${\bigO}(mn)$ time (which is optimal, conditional to $\OVH$) and using only ${\bigO}(\log p)$ bits (which is also optimal for sliding window algorithms, see \cite{GanardiHL16}). 

Complementing the discussion above, one can show that it is possible to construct in linear time, for words $u,w$ and integer $p\in\mathbb N$, a string $w'$ such that deciding whether $u$ is a $\pmas$ of $w'$ is equivalent to deciding whether $u$ is a $p$-subsequence of $w$, so solving $\pSubSeqMatch$ for the input words $u$ and $w$. Hence, the lower bound from \cref{thm:lowerBoundSubseq} carries over,  and the algorithm announced in \cref{thm:algpmas} is optimal (conditional to $\OVH$) from the time complexity point of view. \looseness=-1

Interestingly, the study of absent subsequences was not considered yet for the case of subsequences with gap constraints.

\subsection{Equivalence}\label{sec:equivalence}

The equivalence problem for subsequences $\equiProb$ is to decide, for given strings $v,w\in\Sigma^*$ with $\len v=m$ and $\len w=n$ as well as an integer $k$, whether the sets of subsequences of length at most $k$ of $v$ equals the respective set of $w$, $\Subseq_{\leq k}(v)=\Subseq_{\leq k}(w)$. 

$\equiProb$, and its maximization variant in which one looks for the largest $k$ for which $\equiProb$ with inputs $v,w,k$ is true, were among the most studied problems in relation to subsequences. 
In particular, Hebrard \cite{TCS::Hebrard1991} presented the aforementioned maximization problem as computing a similarity measure between strings and mentions a solution of Simon \cite{SimonUnpublished} for this problem which runs in $O(|\Sigma|nm)$
(the same solution is mentioned in \cite{garelCPM}). Hebrard improves this (see \cite{TCS::Hebrard1991}) in the case when $\Sigma$ is a binary alphabet: given two bitstrings $w$ and $v$, one can find the maximum $k$ for which $\Subseq_{\leq k}(v)=\Subseq_{\leq k}(w)$ in linear time. However, the problem of finding optimal algorithms for both $\equiProb$ and its maximization variant, in the case of general alphabets, was left open in \cite{SimonUnpublished,TCS::Hebrard1991} as the methods used in the latter paper for binary strings did not seem to scale up. In \cite{garelCPM}, Garel approaches the maximization problem and presents an algorithm based on finite automata, running in $O(|\Sigma|n)$, which computes all {\em distinguishing words} $u$ of minimum length,
i.\,e., words which are factors of only one of the words $w$ and $v$ from the problem's statement.
Several further improvements on the aforementioned results were reported in \cite{CrochemoreMT03,DBLP:conf/wia/Tronicek02}. Also, in an extended abstract from 2003 \cite{SimonWords}, Simon presented another algorithm based on finite automata solving this maximization problem, which runs in $O(|\Sigma|n)$, and he conjectures that it can be implemented in $O(|\Sigma|+n)$. Unfortunately, the last claim was only insufficiently substantiated, and obtaining an algorithm with the claimed complexity remained open (in fact, Simon announced that a detailed description of this algorithm will follow shortly, but we were not able to find it in the literature).

Further, in \cite{KufMFCS}, a novel approach to efficiently solving $\equiProb$ was introduced. This idea was to compute, for the two given words $v$ and $w$ and the given number $k$, their shortlex forms: the words which have the same set of subsequences of length at most $k$ as $v$ and $w$, respectively, and are also lexicographically smallest among all words with the respective property.
Clearly, $\Subseq_{\leq k}(v)=\Subseq_{\leq k}(w)$ if and only if the shortlex forms of $v$ and $w$ for $k$ coincide.

The shortlex form of a word $w$ of length $n$ over $\Sigma$ was computed in $O(|\Sigma|n)$ time in \cite{KufMFCS},
so $\equiProb$ was also solved in $O(|\Sigma| n)$. A more efficient implementation of the ideas introduced in~\cite{KufMFCS} was presented in \cite{Barker2020}: the shortlex form of a word of length $n$ over $\Sigma$ can be computed in linear time $O(n)$,
so $\equiProb$ can be solved in optimal linear time.
By binary searching for the smallest $k$ for which $\equiProb$ with inputs $v,w,k$ is true, gives an $O(n\log n)$ time solution for the corresponding optimization problem.\looseness=-1

Later, Gawrychowski et al. \cite{stacs21} solved this optimization problem (finding the maximum $k$ such that $\Subseq_{\leq k}(v)=\Subseq_{\leq k}(w)$) in optimal linear time, as well. However, to achieve this result a novel data structure, the Simon-Tree, was introduced. A node of depth $k$ in the Simon-Tree of a word $w$ corresponds to a maximal interval $[i:j]$ (called $k$-block) such that for all $\ell,\ell'\in [i:j]$ it holds that $\Subseq_k(w[\ell:n])= \Subseq_k(w[\ell':n])$.

\begin{definition}
	The {\em Simon-Tree} $T_w$ associated to the word $w$, with $|w|=n$,
	is an ordered rooted tree.
	The nodes of depth $k$ represent $k-$blocks of $w$, for $0\leq k\leq n$,
	and are defined recursively.\looseness=-1
	\begin{itemize}
		\item The root corresponds to the $0$-block of the word $w$,
		i.\,e., the interval $[1:n]$.
		\item For $k>1$ and for a node $a$ of depth $k-1$,
		which represents a $(k-1)$-block $[i:j]$ with $i<j$,
		the children of $a$ are exactly the blocks of the partition of $[i:j]$ in $k$-blocks,
		ordered decreasingly (right-to-left) by their starting position.
		\item For $k>1$,
		each node of depth $k-1$ which represents a singleton-$(k-1)$-block is a leaf.
	\end{itemize}
\end{definition}  

The Simon-Tree $T_w$ can be constructed in linear time $\bigO(n)$. Furthermore, for words $v, w \in \Sigma^*$ of length $\len v=m$ and $\len w=n$, Gawrychowski et al. (\cite{stacs21}) give a linear time algorithm to connect nodes of the Simon-Trees $T_v,T_{w}$. Two nodes $[i,j]$ in $T_v$ (of depth $k$) and $[i',j']$ in $T_{w}$ (also of depth $k$) become connected if and only if $\Subseq_{\leq k}(v[\ell:m])=\Subseq_{\leq k}(w[\ell':n])$ for all $\ell\in [i:j]$ and $\ell'\in [i',j']$. The maximal $k$ such that $\Subseq_k(v)=\Subseq_k(w)$ now equals the depth of the deepest connected nodes $[1,x]$ in $T_v$ and $[1,y]$ in $T_{w}$. The following result holds.

\begin{theorem}[\cite{stacs21}]
Given two strings $v$ and $w$, with $n=|w|\geq |v|$, the largest $k$ for which $\equiProb$ with input $v,w,k$ is answered positively can be computed in $\bigO(n)$ time.
\end{theorem}

\paragraph{Equivalence and bounded range constraints.}
Considering the problem $\equiProb$ for subsequences occurring within bounded ranges leads again to a surprising intractability result. Once more, for convenience, we give this result for the complement problem, i.\,e., non-equivalence problem $\nequiProb$.
We are given two words $v,w\in\Sigma^*$, and two numbers $k,p\in\mathbb N$, and we want to decide whether $\pSubseq_k(v)\neq \pSubseq_k(w)$. The hardness result obtained in \cite{KKMP22} (and corresponding conditional lower bound) is the following.

\begin{theorem}
	$\nequiProb$ with a bounded range constraint is NP-hard and cannot be solved in subexponential time $2^{\smallO(k)} \poly(k,n,m)$ unless $\ETH$ fails.
\end{theorem}

\paragraph{Equivalence and gap constraints.}
Finally, similarly to the case of subsequences occurring in bounded ranges, for the case of subsequences with gap constraints, one can adapt the results from \cite{Day2022}, related to $\uniProb$, to get the following results.

Firstly, some algorithmic upper bounds.
\begin{theorem}
	(1) The problem $\equiProb$
	for subsequences with length (or reg-len) constraints can be solved in time $\bigO(|\Sigma|^{k} \nz(\gaptuple) \ell)$
	(respectively, $\bigO(|\Sigma|^{k} \states(\gaptuple) \ell)$),
	where $\ell = \max\{|w|, |w'|\}$.\\
	(2) For fixed alphabet $\Sigma$, the problem $\equiProb_{\Sigma}$
	with length (or reg-len) constraints can be solved in time $2^{\bigO(k)} \nz(\gaptuple) \ell$ (respectively, $2^{\bigO(k)} \states(\gaptuple) \ell$), where $\ell = \max\{|w|, |w'|\}$.
\end{theorem}

Secondly, a couple of intractability results, doubled by conditional lower bounds.
\begin{theorem}
	For every fixed alphabet $\Sigma$ with $|\Sigma| \geq 3$, $\nequiProb_{\Sigma}$
	with length constraints is $\npclass$-complete,
	even if all length constraints are $(1, 5)$. Moreover,
	\begin{itemize}
		\item it cannot be solved in subexponential time $2^{\smallO(k)} \poly(|w|, k))$ (unless ETH fails), 
		\item it cannot be solved in time $\bigO(2^{k(1-\epsilon)} \poly(|w|, k))$ (unless SETH fails).
	\end{itemize}
	
	For every fixed alphabet $\Sigma$ with $|\Sigma| = 2$, $\nequiProb_{\Sigma}$
	with length constraints are $\npclass$-complete even if each length constraint is $(0, 0)$ or $(3, 9)$. 
\end{theorem}

\section{Conclusions, Related Problems, and Future Work}

In this paper,
we overviewed a series of recent algorithmic and complexity theoretic results 
related to the matching and analysis problems for (constrained) subsequences. 

An interesting problem related to the study of analysis problems for classical subsequences 
is the \emph{containment problem} 
(denoted $\contProb$ for short), 
which consists in deciding
whether $\Subseq_k(w)\subseteq \Subseq_k(v)$ for given strings $w, v \in \Sigma^*$ and integer $k$. 
This problem can be solved in polynomial time by an automata theoretic approach, as follows. 
We start with our input words $w$ and $v$. 
For $w$ we construct $A_w$, 
the subsequence-automaton \cite{CrochemoreMT03} 
which accepts all the subsequences of $w$. 
Assume $|w|=n$. 
Then, $A_w$ is a deterministic finite automaton, 
which has $n+2$ states $\{0,...,n,n+1\}$. 
The initial state of this automaton is $0$, 
and all states $i$, with $i\in [n]$ are final; 
the state $n+1$ is an error state. 
The transition are defined as follows: 
for all $i\in [n], k\in [n-i]$, and $a\in \Sigma$, 
we have a transition from state $i$ to $i+k$, labelled with $a$, 
if and only if $w[i+k]=a$ and $w[i+1..i+k-1]$ does not contain the letter $a$. 
Moreover, for all $i\in [n]$ and $a\in \Sigma$, 
we have a transition from state $i$ to state $n+1$, labelled with letter $a$, 
if and only if $w[i+1..n]$ does not contain $a$. 
For state $n+1$ we have loop-transitions for all letters $a\in \Sigma$. 
It is straightforward that $A_w$ accepts exactly the non-empty subsequences of $w$ 
and can be constructed in $O(n|\Sigma|)$ time. 
For the word $v$, with $|v|=m$, 
we construct the automaton $A_{v}$, as above, 
and then modify it to obtain the automaton $B_{v}$ by simply making the state $m+1$ the single final state. 
Clearly, $B_{v}$ accepts all strings which are not subsequences of $v$. 
It can be constructed in $O(m|\Sigma|)$ time. 
Now, we can observe that  $\Subseq_k(w)\nsubseteq \Subseq_k(v)$ 
if and only if there exists a word of length at most $k$ accepted by $A_w$ 
which is also accepted by $B_{v}$. 
This can be checked in $O(nm|\Sigma|)$ time 
by simply computing the shortest word in the intersection of the language accepted by $A_w$ with the language accepted by $B_{v}$.
However, it remains an interesting open problem whether a linear time algorithm exists for $\contProb$, 
as it exists for  $\equiProb$. 
 
When considering subsequences with gap constraints, it seems interesting to see if the polynomiality of the matching problem is preserved under adding \emph{gap length equalities} to the gap constraints, i.\,e., constraints of the form $|\gapName_i|=|\gapName_j|$ which are satisfied by an embedding $e$ with respect to $w$ if $|\gap{w}{e}{i}|=|\gap{w}{e}{j}|$. Such length equality constraints (and more complex ones, e.\,g., described by linear inequalities) are of interest in the theory of string solving \cite{amadini2021survey}. Unfortunately, the matching problem becomes immediately NP-hard, as it can be shown (see \cite{Day2022}) by adapting the NP-completeness proof for matching patterns with variables from~\cite{Angluin80}.

Finally, with respect to $\equiProb$, we can consider the following modified setting, both in the classical case and in the case of constrained subsequences (in bounded ranges, or with gap constraints). The idea is to consider the sets of subsequences with multiplicities. We present this here for subsequences with gap constraints, following \cite{Day2022}, but this can be trivially adapted to the case of classical subsequences (see \cite{FreydenbergerGK15}), as well as to the case of subsequences with bounded-range constraints. So, we consider the $\gaptuple$-subsequences of the sets $\subseqSet{\gaptuple}{w}$ with multiplicities. For example, for $w_1 = \ta \tb \tb \ta$ and $w_2 = \ta \tb \ta \tb$, we have $\subseqSet{\gaptuple_2}{w_1} = \subseqSet{\gaptuple_2}{w_2} = \{\ta \ta, \ta \tb, \tb \ta, \tb \tb\}$ with $\gaptuple_2 = (\Sigma^*)$. There is exactly one way of embedding $\ta \ta$ and $\tb \tb$ into both $w_1$ and $w_2$. On the other hand, $\ta \tb$ can be embedded into $w_1$ in two different ways and into $w_2$ in three different ways. More precisely, the sets of $\gaptuple_2$-subsequences of $w_1$ and $w_2$ with multiplicities are $\{(\ta \ta, 1), (\ta \tb, 2), (\tb \ta, 2), (\tb \tb), 1\}$ and $\{(\ta \ta, 1), (\ta \tb, 3), (\tb \ta, 1), (\tb \tb), 1\}$, respectively. So, we can now formalise this setting. For strings $u$ and $v$, and a $(|u|-1)$-tuple $\gaptuple$ of gap constraints, we denote by $\distinctgcsubseq{u}{v}{\gaptuple}$ the number of distinct embeddings $e : |u| \to |v|$ that satisfy $\gaptuple$ and $v \subseq_e u$. For example, $\distinctgcsubseq{\tb \tb \ta \ta}{\tb \ta}{\gaptuple_2} = 4$, as $u[1]u[3] = u[2]u[3] = u[1]u[4] = u[2]u[4] = \tb \ta$. For any $(k-1)$-tuple $\gaptuple$ of gap constraints, we define the function $\parikhK{\cdot}{\gaptuple} \colon \Sigma^* \to \mathbb{N}^{(\Sigma^k)}$ by $\parikhK{w}{\gaptuple}[p] = \distinctgcsubseq{w}{p}{\gaptuple}$ for every $p \in \Sigma^k$. 

The \emph{equivalence problem with multiplicities} is to decide, for a given $(k-1)$-tuple $\gaptuple$ of gap constraints, and strings $w, v \in \Sigma^*$, whether $\parikhK{w}{\gaptuple} = \parikhK{v}{\gaptuple}$. Note that for the case $\gaptuple=(\Sigma^*, \ldots, \Sigma^*)$ this is called the $k$-binomial equivalence, and was studied in the area of combinatorics on words (see, e.\,g., \cite{RigoS15,Rigo19,LeroyRS17a,FreydenbergerGK15}). 
By an automata theoretic approach one can show that the equivalence with multiplicities problem can be decided in polynomial time (in stark contrast to the $\npclass$-completeness of the case without multiplicities). 

The idea, firstly introduced in \cite{FreydenbergerGK15} and then used in \cite{Day2022}, is the following. We first construct, for the first input word $w$ with $\len{w} = n$ and a $(k-1)$-tuple $\gaptuple$ of gap constraints, a non-deterministic finite automaton $A_{w,\gaptuple}$ that accepts exactly the gapped subsequences $p \in \subseqSet{\gaptuple}{w}$ and has exactly $\Psi_{\gaptuple}(w)[p]$ accepting paths labelled with the subsequence $p$ of $w$. Then, we will use the same construction for the second input word $v$, to obtain $A_{v,\gaptuple}$. Finally, we use the algorithm of \cite{Tzeng1992} to test whether $A_{w,\gaptuple}$ and $A_{v,\gaptuple}$ are path equivalent, i.\,e., for each word $p$, the number of accepting paths of $A_{w,\gaptuple}$ labelled with $p$ equals the number of accepting paths of $A_{v,\gaptuple}$ labelled with $p$. If this algorithm returns a positive answer, then we can conclude that $\parikhK{w}{\gaptuple} = \parikhK{v}{\gaptuple}$. Otherwise, we conclude that $\parikhK{w}{\gaptuple} \neq \parikhK{v}{\gaptuple}$.

While the algorithm discussed above related to the \emph{equivalence problem with multiplicities} runs in polynomial time $O(\max\{|w|, |v|\}^4 k^4 + \size(\gaptuple))$ for reg-len constraints, it would be interesting to see if faster algorithms exist. Moreover, the \emph{containment problem with multiplicities} (i.\,e., deciding $\parikhK{w}{\gaptuple}[p] \leq \parikhK{v}{\gaptuple}[p]$ for all $p \in \Sigma^k$) seems to be more difficult. To our knowledge, whether the case of classical subsequences can be solved in polynomial time remains open. 

While the discussion above highlights some clear open problems related to the topics covered in this paper, one could also extend this research by considering, for instance, other classes of constrained subsequences and investigating the matching and analysis problems for those classes as well. 

Several other directions (not covered in this paper), in which the results overviewed here were extended and complemented, are the following. On the one hand, exactly as we did in the case of absent subsequences, one can try to generalise combinatorial and algorithmic properties from factors to subsequences. Such results are reported in, e.g., \cite{abs-2202-07189,BannaiIKKP22} (and the references therein), where the authors are concerned among other with identifying and representing the longest (sub-)periodic subsequences or Lyndon subsequences. On the other hand, several recent works approach topics closely related to Simon's congruence: \cite{CGGLPPS22} focuses on algorithms detecting strings having the same length-k substrings; \cite{KimHKS22} investigates the algorithmic and language theoretic properties of Simon's congruence closure of a string, i.e., the regular set of strings which are $\sim_k$-congruent to a given string, for a given $k$; last, but not least, \cite{KimHK22} solves the string-matching in which one requires finding all factors of a given string $w$ that are $\sim_k$-congruent to another string $v$, for a given $k$. All these works leave many interesting questions open.

We do not claim that this brief overview of related works is exhaustive, but we rather hope it enforces the idea that study of algorithmic properties of subsequences is a vibrant area within combinatorial pattern matching, which already produced some interesting and deep results but also leaves a multitude of challenging open problems. 
\bibliographystyle{eptcs}
\bibliography{references}

\begin{thebibliography}{10}
\providecommand{\bibitemdeclare}[2]{}
\providecommand{\surnamestart}{}
\providecommand{\surnameend}{}
\providecommand{\urlprefix}{Available at }
\providecommand{\url}[1]{\texttt{#1}}
\providecommand{\href}[2]{\texttt{#2}}
\providecommand{\urlalt}[2]{\href{#1}{#2}}
\providecommand{\doi}[1]{doi:\urlalt{https://doi.org/#1}{#1}}
\providecommand{\eprint}[1]{arXiv:\urlalt{https://arxiv.org/abs/#1}{#1}}
\providecommand{\bibinfo}[2]{#2}

\bibitemdeclare{inproceedings}{AbboudEtAl2015}
\bibitem{AbboudEtAl2015}
\bibinfo{author}{Amir \surnamestart Abboud\surnameend}, \bibinfo{author}{Arturs
  \surnamestart Backurs\surnameend} \& \bibinfo{author}{Virginia~Vassilevska
  \surnamestart Williams\surnameend} (\bibinfo{year}{2015}):
  \emph{\bibinfo{title}{Tight Hardness Results for {LCS} and Other Sequence
  Similarity Measures}}.
\newblock In: {\slshape \bibinfo{booktitle}{Proc. {FOCS} 2015}}, pp.
  \bibinfo{pages}{59--78}, \doi{10.1109/FOCS.2015.14}.

\bibitemdeclare{inproceedings}{AbboudEtAl2014}
\bibitem{AbboudEtAl2014}
\bibinfo{author}{Amir \surnamestart Abboud\surnameend},
  \bibinfo{author}{Virginia~Vassilevska \surnamestart Williams\surnameend} \&
  \bibinfo{author}{Oren \surnamestart Weimann\surnameend}
  (\bibinfo{year}{2014}): \emph{\bibinfo{title}{Consequences of Faster
  Alignment of Sequences}}.
\newblock In: {\slshape \bibinfo{booktitle}{Proc. {ICALP} 2014}}, pp.
  \bibinfo{pages}{39--51}, \doi{10.1007/978-3-662-43948-7\_4}.

\bibitemdeclare{article}{amadini2021survey}
\bibitem{amadini2021survey}
\bibinfo{author}{Roberto \surnamestart Amadini\surnameend}
  (\bibinfo{year}{2021}): \emph{\bibinfo{title}{A survey on string constraint
  solving}}.
\newblock {\slshape \bibinfo{journal}{ACM Computing Surveys (CSUR)}}
  \bibinfo{volume}{55}(\bibinfo{number}{1}), pp. \bibinfo{pages}{1--38},
  \doi{10.1145/3484198}.

\bibitemdeclare{article}{Angluin80}
\bibitem{Angluin80}
\bibinfo{author}{Dana \surnamestart Angluin\surnameend} (\bibinfo{year}{1980}):
  \emph{\bibinfo{title}{Finding Patterns Common to a Set of Strings}}.
\newblock {\slshape \bibinfo{journal}{J. Comput. Syst. Sci.}}
  \bibinfo{volume}{21}(\bibinfo{number}{1}), pp. \bibinfo{pages}{46--62},
  \doi{10.1016/0022-0000(80)90041-0}.

\bibitemdeclare{inproceedings}{ArtikisEtAl2017}
\bibitem{ArtikisEtAl2017}
\bibinfo{author}{Alexander \surnamestart Artikis\surnameend},
  \bibinfo{author}{Alessandro \surnamestart Margara\surnameend},
  \bibinfo{author}{Mart{\'{\i}}n \surnamestart Ugarte\surnameend},
  \bibinfo{author}{Stijn \surnamestart Vansummeren\surnameend} \&
  \bibinfo{author}{Matthias \surnamestart Weidlich\surnameend}
  (\bibinfo{year}{2017}): \emph{\bibinfo{title}{Complex Event Recognition
  Languages: Tutorial}}.
\newblock In: {\slshape \bibinfo{booktitle}{Proc. {DEBS} 2017}}, pp.
  \bibinfo{pages}{7--10}, \doi{10.1145/3093742.3095106}.

\bibitemdeclare{article}{DBLP:journals/tcs/Baeza-Yates91}
\bibitem{DBLP:journals/tcs/Baeza-Yates91}
\bibinfo{author}{Ricardo~A. \surnamestart Baeza{-}Yates\surnameend}
  (\bibinfo{year}{1991}): \emph{\bibinfo{title}{Searching Subsequences}}.
\newblock {\slshape \bibinfo{journal}{Theor. Comput. Sci.}}
  \bibinfo{volume}{78}(\bibinfo{number}{2}), pp. \bibinfo{pages}{363--376},
  \doi{10.1016/0304-3975(91)90358-9}.

\bibitemdeclare{inproceedings}{BannaiIKKP22}
\bibitem{BannaiIKKP22}
\bibinfo{author}{Hideo \surnamestart Bannai\surnameend},
  \bibinfo{author}{Tomohiro \surnamestart I\surnameend},
  \bibinfo{author}{Tomasz \surnamestart Kociumaka\surnameend},
  \bibinfo{author}{Dominik \surnamestart K{\"{o}}ppl\surnameend} \&
  \bibinfo{author}{Simon~J. \surnamestart Puglisi\surnameend}
  (\bibinfo{year}{2022}): \emph{\bibinfo{title}{Computing Longest (Common)
  Lyndon Subsequences}}.
\newblock In: {\slshape \bibinfo{booktitle}{Proc. {IWOCA} 2022}}, {\slshape
  \bibinfo{series}{Lecture Notes in Computer Science}} \bibinfo{volume}{13270},
  \bibinfo{publisher}{Springer}, pp. \bibinfo{pages}{128--142},
  \doi{10.1007/978-3-031-06678-8\_10}.
\newblock \bibinfo{note}{{Extended version to appear under the title
  ``Computing Longest Lyndon Subsequences and Longest Common Lyndon
  Subsequences''}}.

\bibitemdeclare{article}{abs-2202-07189}
\bibitem{abs-2202-07189}
\bibinfo{author}{Hideo \surnamestart Bannai\surnameend},
  \bibinfo{author}{Tomohiro \surnamestart I\surnameend} \&
  \bibinfo{author}{Dominik \surnamestart K{\"{o}}ppl\surnameend}
  (\bibinfo{year}{2022}): \emph{\bibinfo{title}{Longest (Sub-)Periodic
  Subsequence}}.
\newblock {\slshape \bibinfo{journal}{CoRR}} \bibinfo{volume}{abs/2202.07189}.
\newblock \eprint{2202.07189}.

\bibitemdeclare{inproceedings}{Barker2020}
\bibitem{Barker2020}
\bibinfo{author}{Laura \surnamestart Barker\surnameend},
  \bibinfo{author}{Pamela \surnamestart Fleischmann\surnameend},
  \bibinfo{author}{Katharina \surnamestart Harwardt\surnameend},
  \bibinfo{author}{Florin \surnamestart Manea\surnameend} \&
  \bibinfo{author}{Dirk \surnamestart Nowotka\surnameend}
  (\bibinfo{year}{2020}): \emph{\bibinfo{title}{Scattered Factor-Universality
  of Words}}.
\newblock In: {\slshape \bibinfo{booktitle}{Proc. DLT 2020}}, {\slshape
  \bibinfo{series}{Lecture Notes in Computer Science}} \bibinfo{volume}{12086},
  pp. \bibinfo{pages}{14--28}, \doi{10.1007/978-3-030-48516-0_2}.

\bibitemdeclare{inproceedings}{CGGLPPS22}
\bibitem{CGGLPPS22}
\bibinfo{author}{Giulia \surnamestart Bernardini\surnameend},
  \bibinfo{author}{Alessio \surnamestart Conte\surnameend},
  \bibinfo{author}{Est{\'{e}}ban \surnamestart Gabory\surnameend},
  \bibinfo{author}{Roberto \surnamestart Grossi\surnameend},
  \bibinfo{author}{Grigorios \surnamestart Loukides\surnameend},
  \bibinfo{author}{Solon~P. \surnamestart Pissis\surnameend},
  \bibinfo{author}{Giulia \surnamestart Punzi\surnameend} \&
  \bibinfo{author}{Michelle \surnamestart Sweering\surnameend}
  (\bibinfo{year}{2022}): \emph{\bibinfo{title}{On Strings Having the Same
  Length- k Substrings}}.
\newblock In: {\slshape \bibinfo{booktitle}{Proc. {CPM} 2022}}, {\slshape
  \bibinfo{series}{LIPIcs}} \bibinfo{volume}{223}, pp.
  \bibinfo{pages}{16:1--16:17}, \doi{10.4230/LIPIcs.CPM.2022.16}.

\bibitemdeclare{article}{BilleEtAl2012}
\bibitem{BilleEtAl2012}
\bibinfo{author}{Philip \surnamestart Bille\surnameend},
  \bibinfo{author}{Inge~Li \surnamestart G{\o}rtz\surnameend},
  \bibinfo{author}{Hjalte~Wedel \surnamestart Vildh{\o}j\surnameend} \&
  \bibinfo{author}{David~Kofoed \surnamestart Wind\surnameend}
  (\bibinfo{year}{2012}): \emph{\bibinfo{title}{String matching with variable
  length gaps}}.
\newblock {\slshape \bibinfo{journal}{Theor. Comput. Sci.}}
  \bibinfo{volume}{443}, pp. \bibinfo{pages}{25--34},
  \doi{10.1016/j.tcs.2012.03.029}.

\bibitemdeclare{inproceedings}{Bringmann2014}
\bibitem{Bringmann2014}
\bibinfo{author}{Karl \surnamestart Bringmann\surnameend}
  (\bibinfo{year}{2014}): \emph{\bibinfo{title}{Why Walking the Dog Takes Time:
  Frechet Distance Has No Strongly Subquadratic Algorithms Unless {SETH}
  Fails}}.
\newblock In: {\slshape \bibinfo{booktitle}{Proc. {FOCS} 2014}}, pp.
  \bibinfo{pages}{661--670}, \doi{10.1109/FOCS.2014.76}.

\bibitemdeclare{inproceedings}{Bringmann2019}
\bibitem{Bringmann2019}
\bibinfo{author}{Karl \surnamestart Bringmann\surnameend}
  (\bibinfo{year}{2019}): \emph{\bibinfo{title}{Fine-Grained Complexity Theory
  (Tutorial)}}.
\newblock In: {\slshape \bibinfo{booktitle}{Proc. {STACS} 2019}}, pp.
  \bibinfo{pages}{4:1--4:7}, \doi{10.4230/LIPIcs.STACS.2019.4}.

\bibitemdeclare{inproceedings}{DBLP:conf/fsttcs/BringmannC18}
\bibitem{DBLP:conf/fsttcs/BringmannC18}
\bibinfo{author}{Karl \surnamestart Bringmann\surnameend} \&
  \bibinfo{author}{Bhaskar~Ray \surnamestart Chaudhury\surnameend}
  (\bibinfo{year}{2018}): \emph{\bibinfo{title}{Sketching, Streaming, and
  Fine-Grained Complexity of (Weighted) {LCS}}}.
\newblock In: {\slshape \bibinfo{booktitle}{Proc. {FSTTCS} 2018}}, {\slshape
  \bibinfo{series}{LIPIcs}} \bibinfo{volume}{122}, pp.
  \bibinfo{pages}{40:1--40:16}, \doi{10.4230/LIPIcs.FSTTCS.2018.40}.

\bibitemdeclare{inproceedings}{BringmannK18}
\bibitem{BringmannK18}
\bibinfo{author}{Karl \surnamestart Bringmann\surnameend} \&
  \bibinfo{author}{Marvin \surnamestart K{\"{u}}nnemann\surnameend}
  (\bibinfo{year}{2018}): \emph{\bibinfo{title}{Multivariate Fine-Grained
  Complexity of Longest Common Subsequence}}.
\newblock In: {\slshape \bibinfo{booktitle}{Proc. {SODA} 2018}}, pp.
  \bibinfo{pages}{1216--1235}, \doi{10.1137/1.9781611975031.79}.

\bibitemdeclare{article}{BussSoltys2014}
\bibitem{BussSoltys2014}
\bibinfo{author}{Sam \surnamestart Buss\surnameend} \& \bibinfo{author}{Michael
  \surnamestart Soltys\surnameend} (\bibinfo{year}{2014}):
  \emph{\bibinfo{title}{Unshuffling a square is {NP}-hard}}.
\newblock {\slshape \bibinfo{journal}{J. Comput. Syst. Sci.}}
  \bibinfo{volume}{80}(\bibinfo{number}{4}), pp. \bibinfo{pages}{766--776},
  \doi{10.1016/j.jcss.2013.11.002}.

\bibitemdeclare{article}{CliffordC07}
\bibitem{CliffordC07}
\bibinfo{author}{Peter \surnamestart Clifford\surnameend} \&
  \bibinfo{author}{Rapha{\"{e}}l \surnamestart Clifford\surnameend}
  (\bibinfo{year}{2007}): \emph{\bibinfo{title}{Simple deterministic wildcard
  matching}}.
\newblock {\slshape \bibinfo{journal}{Inf. Process. Lett.}}
  \bibinfo{volume}{101}(\bibinfo{number}{2}), pp. \bibinfo{pages}{53--54},
  \doi{10.1016/j.ipl.2006.08.002}.

\bibitemdeclare{book}{crochemore}
\bibitem{crochemore}
\bibinfo{author}{Maxime \surnamestart Crochemore\surnameend},
  \bibinfo{author}{Christophe \surnamestart Hancart\surnameend} \&
  \bibinfo{author}{Thierry \surnamestart Lecroq\surnameend}
  (\bibinfo{year}{2007}): \emph{\bibinfo{title}{Algorithms on strings}}.
\newblock \bibinfo{publisher}{Cambridge University Press},
  \doi{10.1017/CBO9780511546853}.

\bibitemdeclare{article}{CrochemoreMT03}
\bibitem{CrochemoreMT03}
\bibinfo{author}{Maxime \surnamestart Crochemore\surnameend},
  \bibinfo{author}{Borivoj \surnamestart Melichar\surnameend} \&
  \bibinfo{author}{Zdenek \surnamestart Tron{\'{\i}}cek\surnameend}
  (\bibinfo{year}{2003}): \emph{\bibinfo{title}{Directed acyclic subsequence
  graph --- Overview}}.
\newblock {\slshape \bibinfo{journal}{J. Discrete Algorithms}}
  \bibinfo{volume}{1}(\bibinfo{number}{3-4}), pp. \bibinfo{pages}{255--280},
  \doi{10.1016/S1570-8667(03)00029-7}.

\bibitemdeclare{inproceedings}{DayFKKMS21}
\bibitem{DayFKKMS21}
\bibinfo{author}{Joel~D. \surnamestart Day\surnameend}, \bibinfo{author}{Pamela
  \surnamestart Fleischmann\surnameend}, \bibinfo{author}{Maria \surnamestart
  Kosche\surnameend}, \bibinfo{author}{Tore \surnamestart Ko{\ss}\surnameend},
  \bibinfo{author}{Florin \surnamestart Manea\surnameend} \&
  \bibinfo{author}{Stefan \surnamestart Siemer\surnameend}
  (\bibinfo{year}{2021}): \emph{\bibinfo{title}{The Edit Distance to
  k-Subsequence Universality}}.
\newblock In: {\slshape \bibinfo{booktitle}{Proc. {STACS} 2021}}, {\slshape
  \bibinfo{series}{LIPIcs}} \bibinfo{volume}{187}, \bibinfo{publisher}{Schloss
  Dagstuhl - Leibniz-Zentrum f{\"{u}}r Informatik}, pp.
  \bibinfo{pages}{25:1--25:19}, \doi{10.4230/LIPIcs.STACS.2021.25}.

\bibitemdeclare{article}{Day2022}
\bibitem{Day2022}
\bibinfo{author}{Joel~D. \surnamestart Day\surnameend}, \bibinfo{author}{Maria
  \surnamestart Kosche\surnameend}, \bibinfo{author}{Florin \surnamestart
  Manea\surnameend} \& \bibinfo{author}{Markus~L. \surnamestart
  Schmid\surnameend} (\bibinfo{year}{2022}): \emph{\bibinfo{title}{Subsequences
  With Gap Constraints: Complexity Bounds for Matching and Analysis Problems}}.
\newblock {\slshape \bibinfo{journal}{CoRR}} \bibinfo{volume}{abs/2206.13896},
  \doi{10.48550/ARXIV.2206.13896}.
\newblock \bibinfo{note}{To appear in the Proceedings of ISAAC 2022.}

\bibitemdeclare{article}{DroubayJP01}
\bibitem{DroubayJP01}
\bibinfo{author}{Xavier \surnamestart Droubay\surnameend},
  \bibinfo{author}{Jacques \surnamestart Justin\surnameend} \&
  \bibinfo{author}{Giuseppe \surnamestart Pirillo\surnameend}
  (\bibinfo{year}{2001}): \emph{\bibinfo{title}{Episturmian words and some
  constructions of de {L}uca and {R}auzy}}.
\newblock {\slshape \bibinfo{journal}{Theor. Comput. Sci.}}
  \bibinfo{volume}{255}(\bibinfo{number}{1-2}), pp. \bibinfo{pages}{539--553},
  \doi{10.1016/S0304-3975(99)00320-5}.

\bibitemdeclare{inproceedings}{KufMFCS}
\bibitem{KufMFCS}
\bibinfo{author}{Lukas \surnamestart Fleischer\surnameend} \&
  \bibinfo{author}{Manfred \surnamestart Kufleitner\surnameend}
  (\bibinfo{year}{2018}): \emph{\bibinfo{title}{Testing {S}imon's congruence}}.
\newblock In: {\slshape \bibinfo{booktitle}{Proc. {MFCS} 2018}}, {\slshape
  \bibinfo{series}{LIPIcs}} \bibinfo{volume}{117}, pp.
  \bibinfo{pages}{62:1--62:13}, \doi{10.4230/LIPIcs.MFCS.2018.62}.

\bibitemdeclare{article}{FleischmannDCFS2022}
\bibitem{FleischmannDCFS2022}
\bibinfo{author}{Pamela \surnamestart Fleischmann\surnameend},
  \bibinfo{author}{Sebastian~Bernhard \surnamestart Germann\surnameend} \&
  \bibinfo{author}{Dirk \surnamestart Nowotka\surnameend}
  (\bibinfo{year}{2021}): \emph{\bibinfo{title}{Scattered Factor Universality -
  The Power of the Remainder}}.
\newblock {\slshape \bibinfo{journal}{CoRR}} \bibinfo{volume}{abs/2104.09063},
  \doi{10.48550/ARXIV.2104.09063}.
\newblock \bibinfo{note}{To appear in Proc. DCFS 2022}.

\bibitemdeclare{article}{Fleischmann2022}
\bibitem{Fleischmann2022}
\bibinfo{author}{Pamela \surnamestart Fleischmann\surnameend},
  \bibinfo{author}{Lukas \surnamestart Haschke\surnameend},
  \bibinfo{author}{Annika \surnamestart Huch\surnameend},
  \bibinfo{author}{Annika \surnamestart Mayrock\surnameend} \&
  \bibinfo{author}{Dirk \surnamestart Nowotka\surnameend}
  (\bibinfo{year}{2022}): \emph{\bibinfo{title}{m-Nearly k-Universal Words -
  Investigating Simon Congruence}}.
\newblock {\slshape \bibinfo{journal}{CoRR}} \bibinfo{volume}{abs/2202.07981},
  \doi{10.48550/ARXIV.2202.07981}.

\bibitemdeclare{article}{FreydenbergerGK15}
\bibitem{FreydenbergerGK15}
\bibinfo{author}{Dominik~D. \surnamestart Freydenberger\surnameend},
  \bibinfo{author}{Pawel \surnamestart Gawrychowski\surnameend},
  \bibinfo{author}{Juhani \surnamestart Karhum{\"{a}}ki\surnameend},
  \bibinfo{author}{Florin \surnamestart Manea\surnameend} \&
  \bibinfo{author}{Wojciech \surnamestart Rytter\surnameend}
  (\bibinfo{year}{2015}): \emph{\bibinfo{title}{Testing $k$-binomial
  equivalence}}.
\newblock {\slshape \bibinfo{journal}{CoRR abs/1509.00622}}, pp.
  \bibinfo{pages}{239--248}, \doi{10.48550/ARXIV.1509.00622}.
\newblock \bibinfo{note}{{\em Multidisciplinary Creativity}, a collection of
  papers dedicated to G. P\u aun 65th birthday}.

\bibitemdeclare{phdthesis}{Ganardi19}
\bibitem{Ganardi19}
\bibinfo{author}{Moses \surnamestart Ganardi\surnameend}
  (\bibinfo{year}{2019}): \emph{\bibinfo{title}{Language recognition in the
  sliding window model}}.
\newblock Ph.D. thesis, \bibinfo{school}{University of Siegen, Germany}.

\bibitemdeclare{inproceedings}{GanardiHKLM18}
\bibitem{GanardiHKLM18}
\bibinfo{author}{Moses \surnamestart Ganardi\surnameend},
  \bibinfo{author}{Danny \surnamestart Hucke\surnameend},
  \bibinfo{author}{Daniel \surnamestart K{\"{o}}nig\surnameend},
  \bibinfo{author}{Markus \surnamestart Lohrey\surnameend} \&
  \bibinfo{author}{Konstantinos \surnamestart Mamouras\surnameend}
  (\bibinfo{year}{2018}): \emph{\bibinfo{title}{Automata Theory on Sliding
  Windows}}.
\newblock In: {\slshape \bibinfo{booktitle}{Proc. {STACS} 2018}}, {\slshape
  \bibinfo{series}{LIPIcs}}~\bibinfo{volume}{96}, \bibinfo{publisher}{Schloss
  Dagstuhl - Leibniz-Zentrum f{\"{u}}r Informatik}, pp.
  \bibinfo{pages}{31:1--31:14}, \doi{10.4230/LIPIcs.STACS.2018.31}.

\bibitemdeclare{inproceedings}{GanardiHL16}
\bibitem{GanardiHL16}
\bibinfo{author}{Moses \surnamestart Ganardi\surnameend},
  \bibinfo{author}{Danny \surnamestart Hucke\surnameend} \&
  \bibinfo{author}{Markus \surnamestart Lohrey\surnameend}
  (\bibinfo{year}{2016}): \emph{\bibinfo{title}{Querying Regular Languages over
  Sliding Windows}}.
\newblock In: {\slshape \bibinfo{booktitle}{Proc. {FSTTCS} 2016}}, {\slshape
  \bibinfo{series}{LIPIcs}}~\bibinfo{volume}{65}, \bibinfo{publisher}{Schloss
  Dagstuhl - Leibniz-Zentrum f{\"{u}}r Informatik}, pp.
  \bibinfo{pages}{18:1--18:14}, \doi{10.4230/LIPIcs.FSTTCS.2016.18}.

\bibitemdeclare{inproceedings}{GanardiHLS19}
\bibitem{GanardiHLS19}
\bibinfo{author}{Moses \surnamestart Ganardi\surnameend},
  \bibinfo{author}{Danny \surnamestart Hucke\surnameend},
  \bibinfo{author}{Markus \surnamestart Lohrey\surnameend} \&
  \bibinfo{author}{Tatiana \surnamestart Starikovskaya\surnameend}
  (\bibinfo{year}{2019}): \emph{\bibinfo{title}{Sliding Window Property Testing
  for Regular Languages}}.
\newblock In: {\slshape \bibinfo{booktitle}{Proc. {ISAAC} 2019}}, {\slshape
  \bibinfo{series}{LIPIcs}} \bibinfo{volume}{149}, \bibinfo{publisher}{Schloss
  Dagstuhl - Leibniz-Zentrum f{\"{u}}r Informatik}, pp.
  \bibinfo{pages}{6:1--6:13}, \doi{10.4230/LIPIcs.ISAAC.2019.6}.

\bibitemdeclare{inproceedings}{garelCPM}
\bibitem{garelCPM}
\bibinfo{author}{Emmanuelle \surnamestart Garel\surnameend}
  (\bibinfo{year}{1993}): \emph{\bibinfo{title}{Minimal Separators of Two
  Words}}.
\newblock In: {\slshape \bibinfo{booktitle}{Proc. CPM 1993}}, {\slshape
  \bibinfo{series}{Lecture Notes in Computer Science}} \bibinfo{volume}{684},
  pp. \bibinfo{pages}{35--53}, \doi{10.1007/BFb0029795}.

\bibitemdeclare{inproceedings}{stacs21}
\bibitem{stacs21}
\bibinfo{author}{Pawel \surnamestart Gawrychowski\surnameend},
  \bibinfo{author}{Maria \surnamestart Kosche\surnameend},
  \bibinfo{author}{Tore \surnamestart Ko{\ss}\surnameend},
  \bibinfo{author}{Florin \surnamestart Manea\surnameend} \&
  \bibinfo{author}{Stefan \surnamestart Siemer\surnameend}
  (\bibinfo{year}{2021}): \emph{\bibinfo{title}{Efficiently Testing Simon's
  Congruence}}.
\newblock In: {\slshape \bibinfo{booktitle}{Proc. {STACS} 2021}}, {\slshape
  \bibinfo{series}{LIPIcs}} \bibinfo{volume}{187}, \bibinfo{publisher}{Schloss
  Dagstuhl - Leibniz-Zentrum f{\"{u}}r Informatik}, pp.
  \bibinfo{pages}{34:1--34:18}, \doi{10.4230/LIPIcs.STACS.2021.34}.

\bibitemdeclare{article}{GiatrakosEtAl2020}
\bibitem{GiatrakosEtAl2020}
\bibinfo{author}{Nikos \surnamestart Giatrakos\surnameend},
  \bibinfo{author}{Elias \surnamestart Alevizos\surnameend},
  \bibinfo{author}{Alexander \surnamestart Artikis\surnameend},
  \bibinfo{author}{Antonios \surnamestart Deligiannakis\surnameend} \&
  \bibinfo{author}{Minos~N. \surnamestart Garofalakis\surnameend}
  (\bibinfo{year}{2020}): \emph{\bibinfo{title}{Complex event recognition in
  the Big Data era: a survey}}.
\newblock {\slshape \bibinfo{journal}{{VLDB} J.}}
  \bibinfo{volume}{29}(\bibinfo{number}{1}), pp. \bibinfo{pages}{313--352},
  \doi{10.1007/s00778-019-00557-w}.

\bibitemdeclare{inproceedings}{HalfonSZ17}
\bibitem{HalfonSZ17}
\bibinfo{author}{Simon \surnamestart Halfon\surnameend},
  \bibinfo{author}{Philippe \surnamestart Schnoebelen\surnameend} \&
  \bibinfo{author}{Georg \surnamestart Zetzsche\surnameend}
  (\bibinfo{year}{2017}): \emph{\bibinfo{title}{Decidability, complexity, and
  expressiveness of first-order logic over the subword ordering}}.
\newblock In: {\slshape \bibinfo{booktitle}{Proc. {LICS} 2017}}, pp.
  \bibinfo{pages}{1--12}, \doi{10.5555/3329995.3330076}.

\bibitemdeclare{article}{TCS::Hebrard1991}
\bibitem{TCS::Hebrard1991}
\bibinfo{author}{Jean-Jacques \surnamestart Hebrard\surnameend}
  (\bibinfo{year}{1991}): \emph{\bibinfo{title}{An algorithm for distinguishing
  efficiently bit-strings by their subsequences}}.
\newblock {\slshape \bibinfo{journal}{Theor. Comput. Sci.}}
  \bibinfo{volume}{82}(\bibinfo{number}{1}), pp. \bibinfo{pages}{35--49},
  \doi{10.1016/0304-3975(91)90170-7}.

\bibitemdeclare{article}{ImpagliazzoPaturi2001}
\bibitem{ImpagliazzoPaturi2001}
\bibinfo{author}{Russell \surnamestart Impagliazzo\surnameend} \&
  \bibinfo{author}{Ramamohan \surnamestart Paturi\surnameend}
  (\bibinfo{year}{2001}): \emph{\bibinfo{title}{On the Complexity of
  $k$-{SAT}}}.
\newblock {\slshape \bibinfo{journal}{J. Comput. Syst. Sci.}}
  \bibinfo{volume}{62}(\bibinfo{number}{2}), pp. \bibinfo{pages}{367--375},
  \doi{10.1006/jcss.2000.1727}.

\bibitemdeclare{article}{ImpagliazzoEtAl2001}
\bibitem{ImpagliazzoEtAl2001}
\bibinfo{author}{Russell \surnamestart Impagliazzo\surnameend},
  \bibinfo{author}{Ramamohan \surnamestart Paturi\surnameend} \&
  \bibinfo{author}{Francis \surnamestart Zane\surnameend}
  (\bibinfo{year}{2001}): \emph{\bibinfo{title}{Which Problems Have Strongly
  Exponential Complexity?}}
\newblock {\slshape \bibinfo{journal}{J. Comput. Syst. Sci.}}
  \bibinfo{volume}{63}(\bibinfo{number}{4}), pp. \bibinfo{pages}{512--530},
  \doi{10.1006/jcss.2001.1774}.

\bibitemdeclare{article}{KarandikarKS15}
\bibitem{KarandikarKS15}
\bibinfo{author}{Prateek \surnamestart Karandikar\surnameend},
  \bibinfo{author}{Manfred \surnamestart Kufleitner\surnameend} \&
  \bibinfo{author}{Philippe \surnamestart Schnoebelen\surnameend}
  (\bibinfo{year}{2015}): \emph{\bibinfo{title}{On the index of {S}imon's
  congruence for piecewise testability}}.
\newblock {\slshape \bibinfo{journal}{Inf. Process. Lett.}}
  \bibinfo{volume}{115}(\bibinfo{number}{4}), pp. \bibinfo{pages}{515--519},
  \doi{10.1016/j.ipl.2014.11.008}.

\bibitemdeclare{inproceedings}{CSLKarandikarS}
\bibitem{CSLKarandikarS}
\bibinfo{author}{Prateek \surnamestart Karandikar\surnameend} \&
  \bibinfo{author}{Philippe \surnamestart Schnoebelen\surnameend}
  (\bibinfo{year}{2016}): \emph{\bibinfo{title}{The Height of
  Piecewise-Testable Languages with Applications in Logical Complexity}}.
\newblock In: {\slshape \bibinfo{booktitle}{Proc. {CSL} 2016}}, {\slshape
  \bibinfo{series}{LIPIcs}}~\bibinfo{volume}{62}, pp.
  \bibinfo{pages}{37:1--37:22}, \doi{10.4230/LIPIcs.CSL.2016.37}.

\bibitemdeclare{article}{journals/lmcs/KarandikarS19}
\bibitem{journals/lmcs/KarandikarS19}
\bibinfo{author}{Prateek \surnamestart Karandikar\surnameend} \&
  \bibinfo{author}{Philippe \surnamestart Schnoebelen\surnameend}
  (\bibinfo{year}{2019}): \emph{\bibinfo{title}{The height of
  piecewise-testable languages and the complexity of the logic of subwords}}.
\newblock {\slshape \bibinfo{journal}{Log. Methods Comput. Sci.}}
  \bibinfo{volume}{15}(\bibinfo{number}{2}), \doi{10.23638/LMCS-15(2:6)2019}.

\bibitemdeclare{article}{KimHK22}
\bibitem{KimHK22}
\bibinfo{author}{Sungmin \surnamestart Kim\surnameend},
  \bibinfo{author}{Yo{-}Sub \surnamestart Han\surnameend} \&
  \bibinfo{author}{Sang{-}Ki \surnamestart Ko\surnameend}
  (\bibinfo{year}{2022}): \emph{\bibinfo{title}{Simon's Congruence Pattern
  Matching}}.
\newblock {\slshape \bibinfo{journal}{To appear in the proceedings of ISAAC
  2022}}.

\bibitemdeclare{inproceedings}{KimHKS22}
\bibitem{KimHKS22}
\bibinfo{author}{Sungmin \surnamestart Kim\surnameend},
  \bibinfo{author}{Yo{-}Sub \surnamestart Han\surnameend},
  \bibinfo{author}{Sang{-}Ki \surnamestart Ko\surnameend} \&
  \bibinfo{author}{Kai \surnamestart Salomaa\surnameend}
  (\bibinfo{year}{2022}): \emph{\bibinfo{title}{On Simon's Congruence Closure
  of a String}}.
\newblock In: {\slshape \bibinfo{booktitle}{Proc. {DCFS} 2022}}, {\slshape
  \bibinfo{series}{Lecture Notes in Computer Science}} \bibinfo{volume}{13439},
  \bibinfo{publisher}{Springer}, pp. \bibinfo{pages}{127--141},
  \doi{10.1007/978-3-031-13257-5\_10}.

\bibitemdeclare{inproceedings}{Markus2022}
\bibitem{Markus2022}
\bibinfo{author}{Sarah \surnamestart Kleest{-}Mei{\ss}ner\surnameend},
  \bibinfo{author}{Rebecca \surnamestart Sattler\surnameend},
  \bibinfo{author}{Markus~L. \surnamestart Schmid\surnameend},
  \bibinfo{author}{Nicole \surnamestart Schweikardt\surnameend} \&
  \bibinfo{author}{Matthias \surnamestart Weidlich\surnameend}
  (\bibinfo{year}{2022}): \emph{\bibinfo{title}{Discovering Event Queries from
  Traces: Laying Foundations for Subsequence-Queries with Wildcards and
  Gap-Size Constraints}}.
\newblock In: {\slshape \bibinfo{booktitle}{Proc. {ICDT} 2022}}, {\slshape
  \bibinfo{series}{LIPIcs}} \bibinfo{volume}{220}, \bibinfo{publisher}{Schloss
  Dagstuhl - Leibniz-Zentrum f{\"{u}}r Informatik}, pp.
  \bibinfo{pages}{18:1--18:21}, \doi{10.4230/LIPIcs.ICDT.2022.18}.

\bibitemdeclare{inproceedings}{Kosche2021}
\bibitem{Kosche2021}
\bibinfo{author}{Maria \surnamestart Kosche\surnameend}, \bibinfo{author}{Tore
  \surnamestart Ko{\ss}\surnameend}, \bibinfo{author}{Florin \surnamestart
  Manea\surnameend} \& \bibinfo{author}{Stefan \surnamestart Siemer\surnameend}
  (\bibinfo{year}{2021}): \emph{\bibinfo{title}{Absent Subsequences in Words}}.
\newblock In: {\slshape \bibinfo{booktitle}{Proc. RP 2021}},
  \bibinfo{publisher}{Springer International Publishing},
  \bibinfo{address}{Cham}, pp. \bibinfo{pages}{115--131},
  \doi{10.1007/978-3-030-89716-1_8}.

\bibitemdeclare{article}{KKMP22}
\bibitem{KKMP22}
\bibinfo{author}{Maria \surnamestart Kosche\surnameend}, \bibinfo{author}{Tore
  \surnamestart Koß\surnameend}, \bibinfo{author}{Florin \surnamestart
  Manea\surnameend} \& \bibinfo{author}{Viktoriya \surnamestart Pak\surnameend}
  (\bibinfo{year}{2022}): \emph{\bibinfo{title}{Subsequences in Bounded Ranges:
  Matching and Analysis Problems}}.
\newblock {\slshape \bibinfo{journal}{CoRR}} \bibinfo{volume}{abs/2207.09201},
  \doi{10.48550/ARXIV.2207.09201}.
\newblock \bibinfo{note}{To appear in the proceedings of RP 2022.}

\bibitemdeclare{inproceedings}{Kuske20}
\bibitem{Kuske20}
\bibinfo{author}{Dietrich \surnamestart Kuske\surnameend}
  (\bibinfo{year}{2020}): \emph{\bibinfo{title}{The Subtrace Order and Counting
  First-Order Logic}}.
\newblock In: {\slshape \bibinfo{booktitle}{Proc. {CSR} 2020}}, {\slshape
  \bibinfo{series}{Lecture Notes in Computer Science}} \bibinfo{volume}{12159},
  pp. \bibinfo{pages}{289--302}, \doi{10.1007/978-3-030-50026-9_21}.

\bibitemdeclare{inproceedings}{KuskeZ19}
\bibitem{KuskeZ19}
\bibinfo{author}{Dietrich \surnamestart Kuske\surnameend} \&
  \bibinfo{author}{Georg \surnamestart Zetzsche\surnameend}
  (\bibinfo{year}{2019}): \emph{\bibinfo{title}{Languages Ordered by the
  Subword Order}}.
\newblock In: {\slshape \bibinfo{booktitle}{Proc. {FOSSACS} 2019}}, {\slshape
  \bibinfo{series}{Lecture Notes in Computer Science}} \bibinfo{volume}{11425},
  pp. \bibinfo{pages}{348--364}, \doi{10.1007/978-3-030-17127-8_20}.

\bibitemdeclare{inproceedings}{Rigo19}
\bibitem{Rigo19}
\bibinfo{author}{Marie \surnamestart Lejeune\surnameend},
  \bibinfo{author}{Julien \surnamestart Leroy\surnameend} \&
  \bibinfo{author}{Michel \surnamestart Rigo\surnameend}
  (\bibinfo{year}{2019}): \emph{\bibinfo{title}{Computing the $k$-binomial
  Complexity of the {T}hue-{M}orse Word}}.
\newblock In: {\slshape \bibinfo{booktitle}{Proc. {DLT} 2019}}, {\slshape
  \bibinfo{series}{Lecture Notes in Computer Science}} \bibinfo{volume}{11647},
  pp. \bibinfo{pages}{278--291}, \doi{10.1007/978-3-030-24886-4_21}.

\bibitemdeclare{article}{LeroyRS17a}
\bibitem{LeroyRS17a}
\bibinfo{author}{Julien \surnamestart Leroy\surnameend},
  \bibinfo{author}{Michel \surnamestart Rigo\surnameend} \&
  \bibinfo{author}{Manon \surnamestart Stipulanti\surnameend}
  (\bibinfo{year}{2017}): \emph{\bibinfo{title}{Generalized {P}ascal triangle
  for binomial coefficients of words}}.
\newblock {\slshape \bibinfo{journal}{Electron. J. Combin.}}
  \bibinfo{volume}{24}(\bibinfo{number}{1.44}), p. \bibinfo{pages}{36 pp.},
  \doi{10.1016/j.aam.2016.04.006}.

\bibitemdeclare{article}{LokshtanovEtAl2011}
\bibitem{LokshtanovEtAl2011}
\bibinfo{author}{Daniel \surnamestart Lokshtanov\surnameend},
  \bibinfo{author}{D{\'{a}}niel \surnamestart Marx\surnameend} \&
  \bibinfo{author}{Saket \surnamestart Saurabh\surnameend}
  (\bibinfo{year}{2011}): \emph{\bibinfo{title}{Lower bounds based on the
  Exponential Time Hypothesis}}.
\newblock {\slshape \bibinfo{journal}{Bull. {EATCS}}} \bibinfo{volume}{105},
  pp. \bibinfo{pages}{41--72}, \doi{10.1007/978-3-319-21275-3_14}.

\bibitemdeclare{article}{LucaGZ08}
\bibitem{LucaGZ08}
\bibinfo{author}{Aldo \surnamestart de~Luca\surnameend}, \bibinfo{author}{Amy
  \surnamestart Glen\surnameend} \& \bibinfo{author}{Luca~Q. \surnamestart
  Zamboni\surnameend} (\bibinfo{year}{2008}): \emph{\bibinfo{title}{Rich,
  Sturmian, and trapezoidal words}}.
\newblock {\slshape \bibinfo{journal}{Theor. Comput. Sci.}}
  \bibinfo{volume}{407}(\bibinfo{number}{1-3}), pp. \bibinfo{pages}{569--573},
  \doi{10.1016/j.tcs.2008.06.009}.

\bibitemdeclare{article}{Maier:1978}
\bibitem{Maier:1978}
\bibinfo{author}{David \surnamestart Maier\surnameend} (\bibinfo{year}{1978}):
  \emph{\bibinfo{title}{The Complexity of Some Problems on Subsequences and
  Supersequences}}.
\newblock {\slshape \bibinfo{journal}{J. ACM}}
  \bibinfo{volume}{25}(\bibinfo{number}{2}), pp. \bibinfo{pages}{322--336},
  \doi{10.1145/322063.322075}.

\bibitemdeclare{article}{Mat04}
\bibitem{Mat04}
\bibinfo{author}{Alexandru \surnamestart Mateescu\surnameend},
  \bibinfo{author}{Arto \surnamestart Salomaa\surnameend} \&
  \bibinfo{author}{Sheng \surnamestart Yu\surnameend} (\bibinfo{year}{2004}):
  \emph{\bibinfo{title}{Subword Histories and {P}arikh Matrices}}.
\newblock {\slshape \bibinfo{journal}{J. Comput. Syst. Sci.}}
  \bibinfo{volume}{68}(\bibinfo{number}{1}), pp. \bibinfo{pages}{1--21},
  \doi{10.1016/j.jcss.2003.04.001}.

\bibitemdeclare{article}{Riddle1979a}
\bibitem{Riddle1979a}
\bibinfo{author}{William~E. \surnamestart Riddle\surnameend}
  (\bibinfo{year}{1979}): \emph{\bibinfo{title}{An Approach to Software System
  Modelling and Analysis}}.
\newblock {\slshape \bibinfo{journal}{Comput. Lang.}}
  \bibinfo{volume}{4}(\bibinfo{number}{1}), pp. \bibinfo{pages}{49--66},
  \doi{10.1016/0096-0551(79)90009-2}.

\bibitemdeclare{article}{RigoS15}
\bibitem{RigoS15}
\bibinfo{author}{Michel \surnamestart Rigo\surnameend} \&
  \bibinfo{author}{Pavel \surnamestart Salimov\surnameend}
  (\bibinfo{year}{2015}): \emph{\bibinfo{title}{Another generalization of
  abelian equivalence: Binomial complexity of infinite words}}.
\newblock {\slshape \bibinfo{journal}{Theor. Comput. Sci.}}
  \bibinfo{volume}{601}, pp. \bibinfo{pages}{47--57},
  \doi{10.1016/j.tcs.2015.07.025}.

\bibitemdeclare{article}{Salomaa05}
\bibitem{Salomaa05}
\bibinfo{author}{Arto \surnamestart Salomaa\surnameend} (\bibinfo{year}{2005}):
  \emph{\bibinfo{title}{Connections Between Subwords and Certain Matrix
  Mappings}}.
\newblock {\slshape \bibinfo{journal}{Theoret. Comput. Sci.}}
  \bibinfo{volume}{340}(\bibinfo{number}{2}), pp. \bibinfo{pages}{188--203},
  \doi{10.1016/j.tcs.2005.03.024}.

\bibitemdeclare{article}{Seki12}
\bibitem{Seki12}
\bibinfo{author}{Shinnosuke \surnamestart Seki\surnameend}
  (\bibinfo{year}{2012}): \emph{\bibinfo{title}{Absoluteness of subword
  inequality is undecidable}}.
\newblock {\slshape \bibinfo{journal}{Theor. Comput. Sci.}}
  \bibinfo{volume}{418}, pp. \bibinfo{pages}{116--120},
  \doi{10.1016/j.tcs.2011.10.017}.

\bibitemdeclare{article}{Shaw1978}
\bibitem{Shaw1978}
\bibinfo{author}{Alan~C. \surnamestart Shaw\surnameend} (\bibinfo{year}{1978}):
  \emph{\bibinfo{title}{Software Descriptions with Flow Expressions}}.
\newblock {\slshape \bibinfo{journal}{{IEEE} Trans. Software Eng.}}
  \bibinfo{volume}{4}(\bibinfo{number}{3}), pp. \bibinfo{pages}{242--254},
  \doi{10.1109/TSE.1978.231501}.

\bibitemdeclare{article}{SimonUnpublished}
\bibitem{SimonUnpublished}
\bibinfo{author}{Imre \surnamestart Simon\surnameend}: \emph{\bibinfo{title}{An
  Algorithm to Distinguish Words efficiently by their Subwords}}.
\newblock {\slshape \bibinfo{journal}{unpublished}}.

\bibitemdeclare{phdthesis}{simonPhD}
\bibitem{simonPhD}
\bibinfo{author}{Imre \surnamestart Simon\surnameend} (\bibinfo{year}{1972}):
  \emph{\bibinfo{title}{Hierarchies of events with dot-depth one}}.
\newblock Ph.D. thesis.

\bibitemdeclare{inproceedings}{Simon72}
\bibitem{Simon72}
\bibinfo{author}{Imre \surnamestart Simon\surnameend} (\bibinfo{year}{1975}):
  \emph{\bibinfo{title}{Piecewise testable events}}.
\newblock In: {\slshape \bibinfo{booktitle}{Autom.\ Theor.\ Form.\ Lang., 2nd
  GI Conf.}}, {\slshape \bibinfo{series}{LNCS}}~\bibinfo{volume}{33}, pp.
  \bibinfo{pages}{214--222}, \doi{10.1007/3-540-07407-4_23}.

\bibitemdeclare{inproceedings}{SimonWords}
\bibitem{SimonWords}
\bibinfo{author}{Imre \surnamestart Simon\surnameend} (\bibinfo{year}{2003}):
  \emph{\bibinfo{title}{Words distinguished by their subwords (extended
  Abstract)}}.
\newblock In: {\slshape \bibinfo{booktitle}{Proc. {WORDS} 2003}}, {\slshape
  \bibinfo{series}{TUCS General Publication}}~\bibinfo{volume}{27}, pp.
  \bibinfo{pages}{6--13}.

\bibitemdeclare{inproceedings}{DBLP:conf/wia/Tronicek02}
\bibitem{DBLP:conf/wia/Tronicek02}
\bibinfo{author}{Zdenek \surnamestart Tron{\'{\i}}cek\surnameend}
  (\bibinfo{year}{2002}): \emph{\bibinfo{title}{Common Subsequence Automaton}}.
\newblock In: {\slshape \bibinfo{booktitle}{Proc. {CIAA} 2002 (Revised
  Papers)}}, {\slshape \bibinfo{series}{Lecture Notes in Computer Science}}
  \bibinfo{volume}{2608}, pp. \bibinfo{pages}{270--275},
  \doi{10.1007/3-540-44977-9_28}.

\bibitemdeclare{article}{Tzeng1992}
\bibitem{Tzeng1992}
\bibinfo{author}{Wen{-}Guey \surnamestart Tzeng\surnameend}
  (\bibinfo{year}{1992}): \emph{\bibinfo{title}{A Polynomial-Time Algorithm for
  the Equivalence of Probabilistic Automata}}.
\newblock {\slshape \bibinfo{journal}{{SIAM} J. Comput.}}
  \bibinfo{volume}{21}(\bibinfo{number}{2}), pp. \bibinfo{pages}{216--227},
  \doi{10.1137/0221017}.

\bibitemdeclare{inproceedings}{Williams2015}
\bibitem{Williams2015}
\bibinfo{author}{Virginia~Vassilevska \surnamestart Williams\surnameend}
  (\bibinfo{year}{2015}): \emph{\bibinfo{title}{Hardness of Easy Problems:
  Basing Hardness on Popular Conjectures such as the Strong Exponential Time
  Hypothesis (Invited Talk)}}.
\newblock In: {\slshape \bibinfo{booktitle}{Proc. {IPEC} 2015}}, pp.
  \bibinfo{pages}{17--29}, \doi{10.4230/LIPIcs.IPEC.2015.17}.

\bibitemdeclare{inproceedings}{Zetzsche16}
\bibitem{Zetzsche16}
\bibinfo{author}{Georg \surnamestart Zetzsche\surnameend}
  (\bibinfo{year}{2016}): \emph{\bibinfo{title}{The Complexity of Downward
  Closure Comparisons}}.
\newblock In: {\slshape \bibinfo{booktitle}{Proc. {ICALP} 2016}}, {\slshape
  \bibinfo{series}{LIPIcs}}~\bibinfo{volume}{55}, pp.
  \bibinfo{pages}{123:1--123:14}, \doi{10.4230/LIPIcs.ICALP.2016.123}.

\bibitemdeclare{inproceedings}{ZhangEtAl2014}
\bibitem{ZhangEtAl2014}
\bibinfo{author}{Haopeng \surnamestart Zhang\surnameend},
  \bibinfo{author}{Yanlei \surnamestart Diao\surnameend} \&
  \bibinfo{author}{Neil \surnamestart Immerman\surnameend}
  (\bibinfo{year}{2014}): \emph{\bibinfo{title}{On complexity and optimization
  of expensive queries in complex event processing}}.
\newblock In: {\slshape \bibinfo{booktitle}{Proc. {SIGMOD} 2014}}, pp.
  \bibinfo{pages}{217--228}, \doi{10.1145/2588555.2593671}.

\end{thebibliography}

\end{document}